\documentclass[review]{elsarticle}
\usepackage{framed}
\usepackage{multicol}
\usepackage{subcaption}
\usepackage{graphicx}
\usepackage{amsmath}
\usepackage{color,soul}
\usepackage{soul}
\usepackage{xcolor}
\usepackage{rotating}
\usepackage{comment}

\usepackage{nomencl}
\makenomenclature

\usepackage{lineno,hyperref}

\usepackage{geometry}
\makeatletter
\def\ps@pprintTitle{%
	\let\@oddhead\@empty
	\let\@evenhead\@empty
	\def\@oddfoot{}%
	\let\@evenfoot\@oddfoot}
\makeatother

\begin{document}

\begin{frontmatter}

\title{On the heat transfer coefficient jump in tilted single-phase natural circulation systems}

\author{Akhil Dass, Sateesh Gedupudi \footnote{Corresponding author. Tel.: +91 44 2257 4721, Email: sateeshg@iitm.ac.in}}
\address{Heat Transfer and Thermal Power Laboratory, Department of Mechanical Engineering, IIT Madras, Chennai 600036, India}

\begin{abstract}
The present study investigates the effect of inclination on the buoyancy driven systems. A cavity, Natural Circulation Loop (NCL) and Coupled Natural Circulation Loop (CNCL) have been used to represent the possible classes of buoyancy driven systems. There are extensive studies on the cavity systems which exhibit a heat transfer coefficient jump upon the change in the inclination as the system moves from one steady state to another. The primary objective of the present study is to investigate if a similar behavior occurs in natural circulation systems such as an NCL and a CNCL, using a 2-D CFD study.
The study identifies a combination of flow pattern rearrangement and flow direction reversal as the phenomena responsible for the jump in the heat transfer coefficient. A 3-D CFD study is also conducted to demonstrate the universality of the phenomena. The CFD study is validated thoroughly with the numerical and experimental data available in the literature. A parametric study is also conducted to examine the effects of the aspect ratio, Rayleigh number, Prandtl number and the inclination with the plane of the NCL system.
\end{abstract}


\end{frontmatter}

\nolinenumbers

\section{Introduction}

Buoyancy driven systems such as thermally excited cavity (referred to simply as the cavity from hereon), a Natural Circulation Loop (NCL) and Coupled Natural Circulation Loop (CNCL) systems have umpteen applications in various engineering domains primarily because of their passive nature (i.e., they do not require active control) and the lack of moving parts which ensures a longer life span of operation and lower maintenance.

Apart from the aforementioned advantages of the system, the systems also have generated a lot of research interest due to their interesting properties such as:
\begin{enumerate}
	\item The existence of multiple steady state solutions \cite{Yang1988},\cite{Acosta1987}.
	\item Exhibiting chaotic behavior at higher heat loads \cite{Yang1988},\cite{Fichera2003}.
	\item Exhibiting hysteresis behavior \cite{Yang1988},\cite{Acosta1987}.  
\end{enumerate}

\subsection{Literature on cavity systems}

Numerous studies have been conducted on the heat transfer in cavities and explored in detail the physics of the process. A detailed review of the work on convection in cavities prior to 1972 was carried out by Ostrach (1972) \cite{Ostrach1972}. As the focus of the work done in the current study is to examine the effect of inclination on buoyancy driven systems, the literature pertaining to inclination studies in cavity systems is reviewed and has been classified into:

\begin{enumerate}[(a)]
	\item \textit{Experimental study}: The existence of heat transfer coefficient jump or discontinuity in cavity system w.r.t inclination has been experimentally reported by Ozoe et al. \cite{Ozoe1974} in 1974. A CFD study was also conducted and a good match between experimental and CFD predictions were reported, but the discontinuity was not captured in the CFD study. In 1975 Ozoe et al.\cite{Ozoe1975} conducted further parametric study on the inclined cavity system to analyse the effect of aspect ratio and effect of variation of Rayleigh number. Symons and Peck \cite{Symons1984} also conducted experiments in 1979 on cavity systems where they examined the effect of inclination along the transverse and longitudinal directions and observed heat transfer coefficient jump or discontinuity for both the cases. The next experiment on the influence of inclination on the cavity system was conducted by Hamady et al. \cite{Hamady1989} in 1989 where both numerical and experimental study was conducted and a good match was observed in the flow patterns predicted by the numerical study with experimental data. The effect of inclination on the heat transfer of the cavity system was also studied and a heat transfer coefficient discontinuity was observed. Two other experimental studies which consider the effect of inclination on cavity systems but have not observed any heat transfer coefficient jump or discontinuity are Bairi (2008) \cite{Bairi2008} and Madanan and Goldstein (2019)\cite{Madanan2019}. The step-sizes of the inclination considered in their study were $45^\circ$ and $30^\circ$ respectively, which are too large to observe any heat transfer coefficient jump.
	
	\item \textit{CFD study}: The CFD work may further be classified into 2-D and 3-D CFD studies.
	\begin{enumerate}[(i)]
		\item \textit{2-D CFD study}: Kuyper et al. \cite{Kuyper1993} in 1993 studied the effect of inclination on cavity systems for both laminar and turbulent flow regimes and observed the heat transfer coefficient jump and hysteresis effect. A more detailed 2-D study of cavity systems was carried out by Soong et al. \cite{Soong1996} in 1996. The effect of different aspect ratios on the inclined cavity systems and the mechanism through which the heat transfer coefficient jump/discontinuity occurs is examined. For cavity systems with aspect ratios greater than one, the flow pattern rearrangement from unicellular to multicellular or vice-versa has been observed to cause the heat transfer coefficient jump. A good agreement with experimental data was reported after accounting for the possible deviations in the boundary conditions. Cianfrini et al. \cite{Cianfrini2005} in 2005 observed the same phenomena in cavity systems with inclination and proposed a piece-wise correlation for the Nusselt number as a function of the inclination angle taking into account the hysteresis and the heat transfer coefficient jump. To get a better understanding of the behaviour of cavity systems a detailed parametric study was conducted by Khezzar et al. \cite{Khezzar2012} in 2012 and it was observed that for few cases the hysteresis behaviour was not present. Most recent 2-D study on cavity systems was conducted by Chang \cite{Chang2014} in 2014. The work presents a detailed parametric study considering various aspect ratios and proposed correlations for the heat transfer for low aspect ratio systems.
		
		\item \textit{3-D CFD study}: The first 3-D CFD study of cavity systems was carried out by Yang et al. \cite{Yang1987} in 1987. It was observed that with inclination the cavity systems exhibited flow pattern transition from multicellular to unicellular for large aspect ratio systems. To identify the regions where the flow transition occurs, a stability map of the system was obtained by Yang \cite{Yang1988} in 1988. An extensive study of transitions and bifurcations of laminar buoyant flows in cavity systems has been conducted. The stability maps of cavity systems were obtained employing linear stability analysis and the inclination effect on the flow patterns of the cavity systems was also examined.
	\end{enumerate}
	
\end{enumerate}

\subsection{Literature on rectangular natural circulation systems}

The literature available on the tilted rectangular natural circulation systems is reviewed and presented as follows:

\begin{enumerate}[(a)]
	\item \textit{Experimental study}: Acosta et al. \cite{Acosta1987} in 1987 were the first to study experimentally the effect of inclination on NCL system. They considered a square NCL system with heat flux condition at the bottom and constant temperature boundary condition at the top leg, respectively. The hysteresis phenomena w.r.t. the inclination angle was observed and the study also verified the existence of multiple steady state solutions in NCL systems. The next study on the effect of inclination on NCL system was studied in 2005 by Misale et al. \cite{Misale2005} for two different fluids and the inclination effect was used to control the net gravity on the system and thereby to control the system stability. This was followed by experiments on mini NCL systems, where the system transience for different inclinations was reported by Misale et al \cite{Misale2007} and Garibaldi and Misale \cite{Garibaldi2008}. The most recent study on the effect of inclination on NCL systems was conducted by Tian et al. \cite{Tian2017} and Zhu et al. \cite{Zhu2017}. It was noted that the heat transfer coefficient is significantly affected by the inclination and both clockwise and anti-clockwise inclinations were considered for the study. Vijayan et al. (2007) \cite{vijayan2007steady} observed that introducing a small inclination to the NCL system changed the transient dynamics of the system.  
	
	\item  \textit{CFD study} : Ramos et al. \cite{Ramos1990} in 1990 conducted a 2-D CFD study of a rectangular NCL system on the effect of inclination on the streamline pattern rearrangement. It was observed that the streamlines rearrangement happened after a certain inclination angle which was termed as the critical angle. Bouali et al. \cite{Bouali2006} in 2006 conducted a 2-D CFD study to investigate the effect of radiation and natural convection  in cavity and NCLs. The effect of inclination was also examined and was reported to have a strong influence on the isotherm patterns. Recently 3-D CFD studies were performed by Basu et al. (2013) \cite{Basu2013}  and Krishnani and Basu (2017) \cite{Krishnani2017} where the effect of inclination on the transient behaviour of NCL systems was investigated and tilting the NCL was observed to have a stabilizing effect on the system transience.
	
\end{enumerate}

\subsection{Objective and scope of the present study}

The advantages and special characteristics of buoyancy driven system may also lead to interesting heat transfer characteristics of such systems. Soong et al. (1996) \cite{Soong1996} have already demonstrated that there is a jump in heat transfer coefficient which occurs due to flow mode transition in a tilted cavity of aspect ratio 4. These observations motivates a natural question as to whether such phenomena is restricted to cavity or does it extend to other buoyancy driven systems as well? To address this question the following objectives are set in the current study:

\begin{enumerate}
	\item Determine if tilted NCL and CNCL systems exhibit heat transfer coefficient jump.
	\item To identify the mechanism of the heat transfer coefficient jump in tilted NCL and CNCL systems if it occurs.
	\item To identify if there is a hysteresis effect in tilted NCL and CNCL systems.
	\item To study the influence of different parameters on the behaviour of tilted NCL and CNCL systems.
\end{enumerate}

\begin{figure}[!htb]
	\centering
	\includegraphics[width=0.6\linewidth]{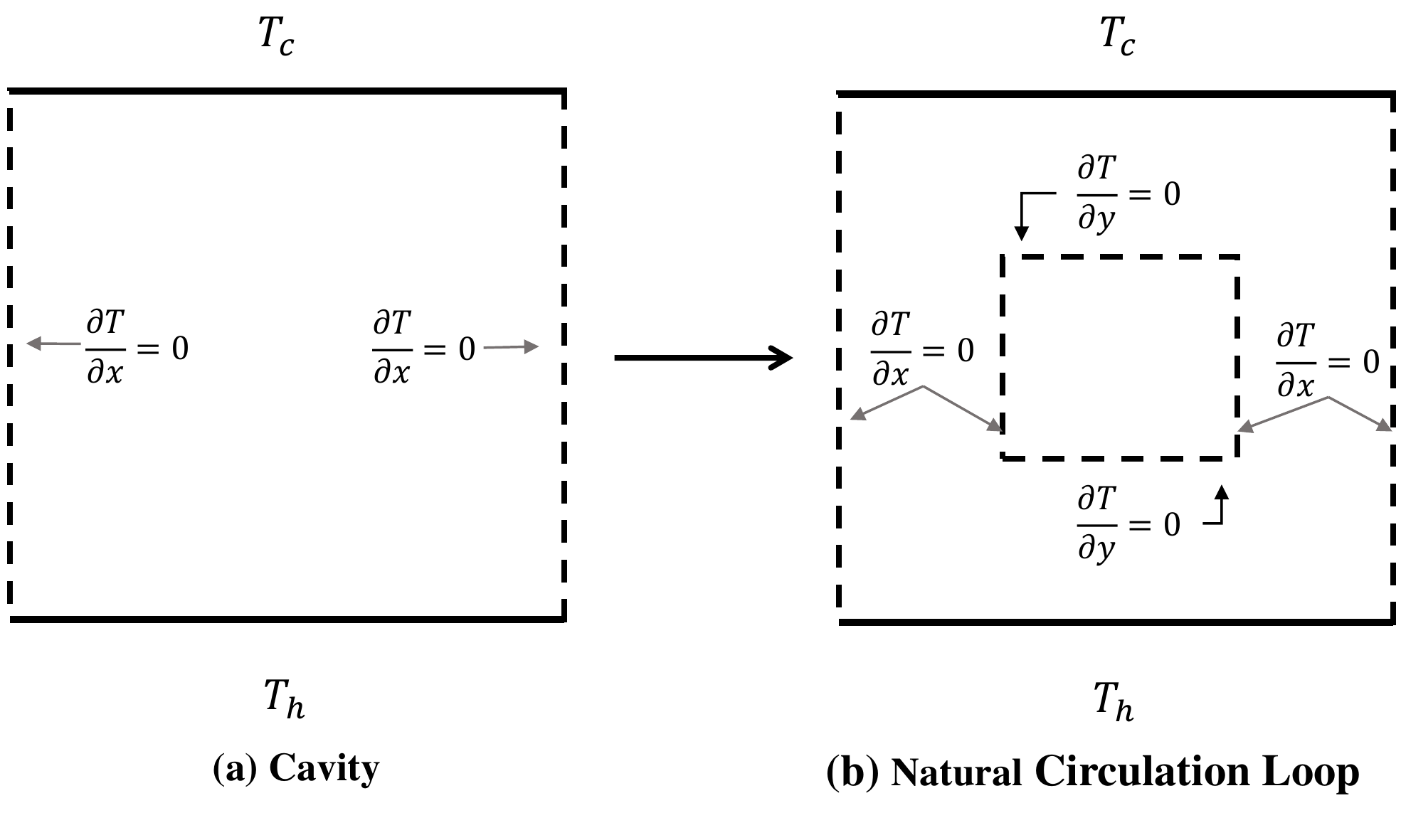}
	\caption{Demonstrating the link between a cavity and natural circulation loop. The dashed lines are used to indicate thermally insulated walls.}
	\label{fig:linkcavityncl}
\end{figure}

 A cavity is a fluid filled enclosure, which is generally heated at the bottom and cooled at the top. The thermal excitation provided generates buoyancy force which circulates the fluid in the entire volume and transfers heat from the hot wall to the cold wall. An NCL can be considered as a fluid filled conduit which is closed from end to end and is driven by buoyancy when subjected to external thermal stimulus. The sole difference between a cavity and an NCL system is in their geometry. Figure \ref{fig:linkcavityncl} clearly demonstrates that an NCL system can be obtained from a 2-D cavity by just introducing another thermally insulated cavity in the primary cavity system. 

The NCL is generally used as a basic representative element of systems which are actually used in practice. In practical applications, the byouancy driven systems such as the Passive Residual Heat Removal System (PRHRS) employ multiple NCLs which are thermally coupled. Thus to generalize the present study it is also necessary to incorporate the effect of thermal coupling on NCL system dynamics. This is done by considering a CNCL system which comprises of two thermally linked NCLs at the common heat exchange section. The recent paper by Dass and Gedupudi (2019) \cite{Dass2019} explores the link between PRHRS and NCL system and presents a CNCL as an appropriate and simplified basic representative of PRHRS. Thus, the effect of tilting the CNCL on the heat transfer at the coupled section is also explored in the current study.

The paper begins with a 2D CFD study of the buoyancy driven systems namely: cavity, NCL and CNCL. The geometry, meshing, governing equations and the appropriate settings and discretizations used are described. A mesh independence study is undertaken followed by a thorough validation with the literature data indicating the reliability of the CFD study.  The present work makes a comparison of the 3-D CFD results with  the experimental data available in the literature, as a validation study.

\section{Contributions of the present study}
\begin{enumerate}
	\item The current work investigates the occurrence of heat transfer coefficient jump and hysteresis in inclined NCL and CNCL systems and presents a common mechanism to explain the observed heat transfer coefficient variation with respect to the inclination of such systems.
	\item A detailed 3-D CFD validation with experimental data is conducted to demonstrate the reliability of the present study and establish the validity of the mechanism of the heat transfer coefficient jump.
	\item A thorough 2-D parametric study on the effect of variation of various non-dimensional numbers on the heat transfer coefficient jump in NCL systems is presented.
	\item The effect of curvature of the bends and the heater cooler arrangement of the NCL is also studied. 
	\item An assessment of the previous literature on inclined NCL systems is conducted and the possible reasons for overlooking of the heat transfer coefficient jump in such  systems are discussed.
\end{enumerate}

\section{Methodology}

To examine the influence of inclination of buoyancy driven systems a thorough and systematic 2-D CFD study is undertaken. The buoyancy driven systems which are considered for the present study are: (a) Cavity (b) Natural Circulation Loop (NCL) and (c) Coupled Natural Circulation Loop (CNCL). The CFD study is conducted employing ANSYS Fluent 16.1 software.

\subsection{Geometry}

\begin{figure}[!htb]
	\centering
	\includegraphics[width=\linewidth]{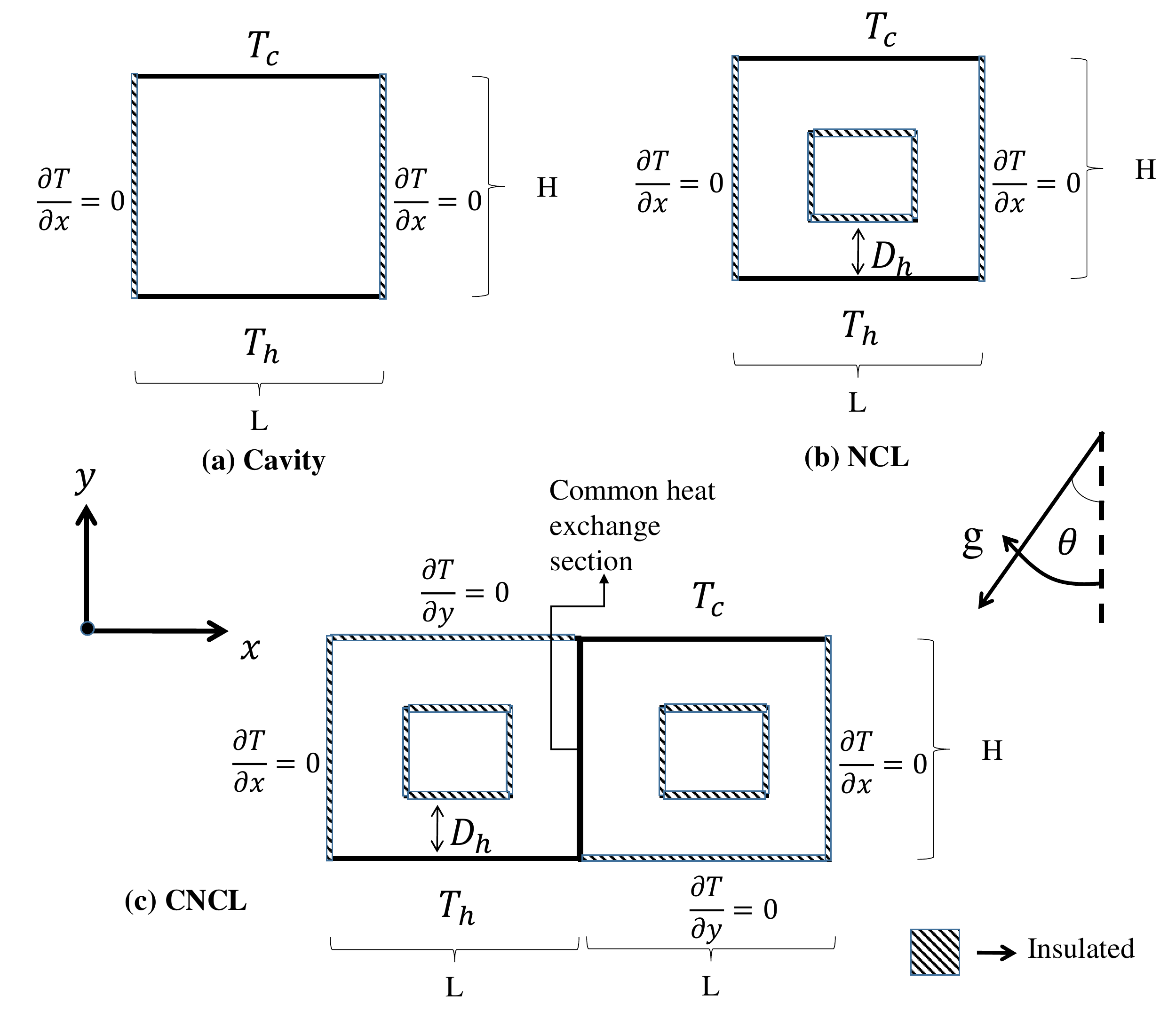}
	\caption{Buoyancy driven systems considered for the present study: (a) Cavity, (b) Natural Circulation Loop (NCL), (c) Coupled Natural Circulation Loop (CNCL). }
	\label{fig:Geometry}
\end{figure}

The buoyancy driven systems which are considered for the present study are represented in Figure \ref{fig:Geometry}. The systems are thermally excited by constant temperature conditions at the top and bottom of the geometry with rest of the loop being thermally insulated (with the exception of the common heat exchange section of the CNCL system). The bottom sections are maintained at a temperature $T_h$ and the top sections are maintained at a temperature $T_c$ with $T_h>T_c$ to obtain convection within the system. The height of the system is denoted by $H$, width by $L$ and $D_h$ denotes the hydraulic diameter of the NCL and CNCL systems. The gravity of the system subtends an angle $\theta$ with respect to the vertical, which has the equivalent effect of tilting the buoyancy systems by the same angle w.r.t. the horizontal. The system makes an angle $\theta$ with horizontal in the anti-clockwise direction or the system is held fixed and the gravity direction is varied in the clockwise direction. For the present study the angle of inclination is varied from $\theta=0^\circ$ to $\theta=90^\circ$ and the influence on the heat transfer characteristics is studied.

\subsection{Governing equations}

\subsubsection{Dimensional governing equations}

The steady state 2-D governing equations of continuity, momentum and energy used for the present CFD study after employing the Boussinesq approximation are:

\begin{equation}
\frac{\partial u}{\partial x} + \frac{\partial v}{\partial y}=0
\end{equation}

\begin{equation}
u\frac{\partial u}{\partial x} + v\frac{\partial u}{\partial y}= \nu \nabla^2 u - \frac{1}{\rho}\frac{\partial p}{\partial x} + g\beta(T-T_0) sin(\theta)
\end{equation}

\begin{equation}
u\frac{\partial v}{\partial x} + v\frac{\partial v}{\partial y}= \nu \nabla^2 v - \frac{1}{\rho}\frac{\partial p}{\partial y} + g\beta(T-T_0) cos(\theta)
\end{equation}

\begin{equation}
u\frac{\partial T}{\partial x} + v\frac{\partial T}{\partial y}= \alpha \nabla^2 T 
\end{equation}

\subsubsection{Non-dimensional governing equations}

To identify the basic non-dimensional numbers which characterize the buoyancy driven systems, the governing equations are non-dimensionalised using the following dimensionless variables:

\begin{equation}
X=\frac{x}{H} \;\;\; Y=\frac{y}{H}
\end{equation}

\begin{equation}
 U=\frac{uH}{\alpha}\;\;\; V=\frac{vH}{\alpha}\;\;\; 
\end{equation}

\begin{equation}
P=\frac{pH^2}{\rho \alpha^2}\;\;\; \phi=\frac{T-T_0}{T_h-T_c}
\end{equation}

Substituting the dimensionless variables listed in equation (5) into equations (1)-(4) we obtain 

\begin{equation}
\frac{\partial U}{\partial X} + \frac{\partial V}{\partial Y}=0
\end{equation}

\begin{equation}
 U\frac{\partial U}{\partial X} + V\frac{\partial U}{\partial Y}= Pr \nabla^2 U - \frac{\partial P}{\partial X} + Ra\; Pr \; \phi \; sin(\theta) 
\end{equation}

\begin{equation}
 U\frac{\partial V}{\partial X} + V\frac{\partial V}{\partial Y}= Pr \nabla^2 V - \frac{\partial P}{\partial Y} + Ra\; Pr \; \phi \; cos(\theta) 
\end{equation}

\begin{equation}
U\frac{\partial \phi}{\partial X} + V\frac{\partial \phi}{\partial Y}= \nabla^2 \phi 
\end{equation}

where

 \begin{equation}
 Ra=\frac{g \beta (T_h-T_c)H^3}{\alpha \nu}
 \end{equation}

\begin{equation}
Pr=\frac{\nu}{\alpha}
\end{equation}

From the aforementioned equations it can be concluded that $Ra$ and $Pr$ are the non-dimensional numbers which influence the dynamics of all the buoyancy driven systems considered in the present study. 

\subsection{Mesh details}

\begin{figure}[!htb]
	\centering
	\includegraphics[width=\linewidth]{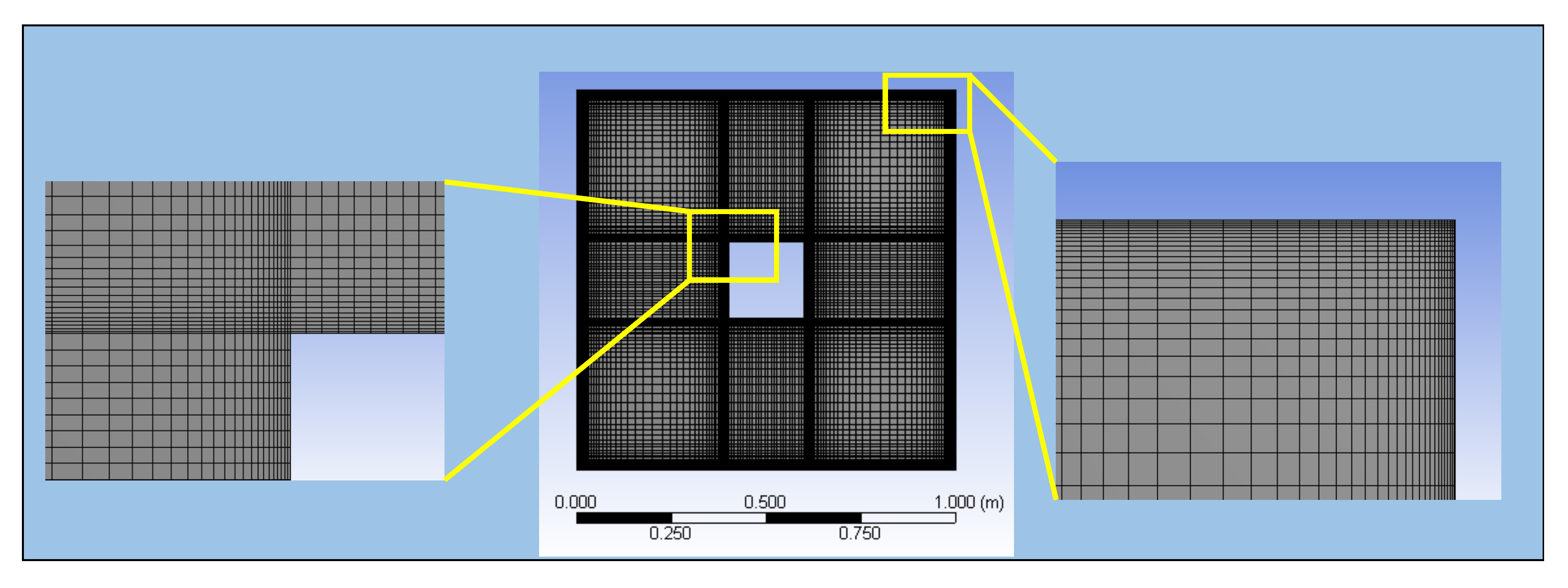}
	\caption{Mesh used to study the NCL system.}
	\label{fig:nclmesh}
\end{figure}

A structured mesh is used for the 2-D CFD study with inflation provided at the walls to capture the boundary layer effects. The mesh used in the present study for an NCL system is represented in Figure \ref{fig:nclmesh}. Similar mesh is utilized for the study of cavity and CNCL systems as well. 

\subsection{Case setup}

A steady state pressure based solver is used for the present study. The flow is laminar, and the Boussinesq approximation is used to couple the velocity and temperature fields. The SIMPLE scheme is used for pressure-velocity coupling. Second order spatial discretization is used for pressure and a second order upwind based spatial discretization is utilized for momentum and energy. The  scaled residuals of $10^{-6}$ is chosen as the criteria for covergence. For the present 2-D study the $Ra$ value is kept fixed at a constant magnitude and the $Pr$ value is taken as 0.7, which corresponds to air as the working fluid. The Nusselt number ($Nu_h$) and Reynolds numbers ($Re$) are calculated along with the streamline profiles to understand the effects of inclination on the heat transfer characteristics of buoyancy driven systems. 

The buoyancy driven systems are prone to hysteresis and to demonstrate this phenomena in all the considered systems, the system with zero inclination is considered as the base case for $\theta$ considered in the ascending order from $0^\circ$ to $90^\circ$. For all $\theta>0$, the previous steady state solution is chosen as the initial condition. For example, if we are increasing $\theta$ in steps of $5^\circ$, then for $\theta=5^\circ$ the initial condition is the steady state solution at $\theta=0^\circ$, and for $\theta=10^\circ$ the initial condition is the steady state solution at $\theta=5^\circ$ or $\theta=0^\circ$ and so on till $\theta=90^\circ$. And for obtaining the effect of inclination for the decreasing angle of $\theta$, $\theta=90^\circ$ is considered as the base case. The initial condition for an inclination $\theta$ is any previous steady state solution for any higher angle in the direction of $90^\circ$ to $\theta$. The initial conditions used for obtaining the steady state solutions for the base cases at $\theta=0^\circ$ and $\theta=90^\circ$ are:

\begin{equation}
U=0, \; V=0, \; \phi=0 \; (\mbox{For the entire domain})
\end{equation}

\subsection{Mesh independence test}

\begin{figure}[!htb]
	\centering
	\begin{subfigure}[b]{0.49\textwidth}
		\includegraphics[width=1\linewidth]{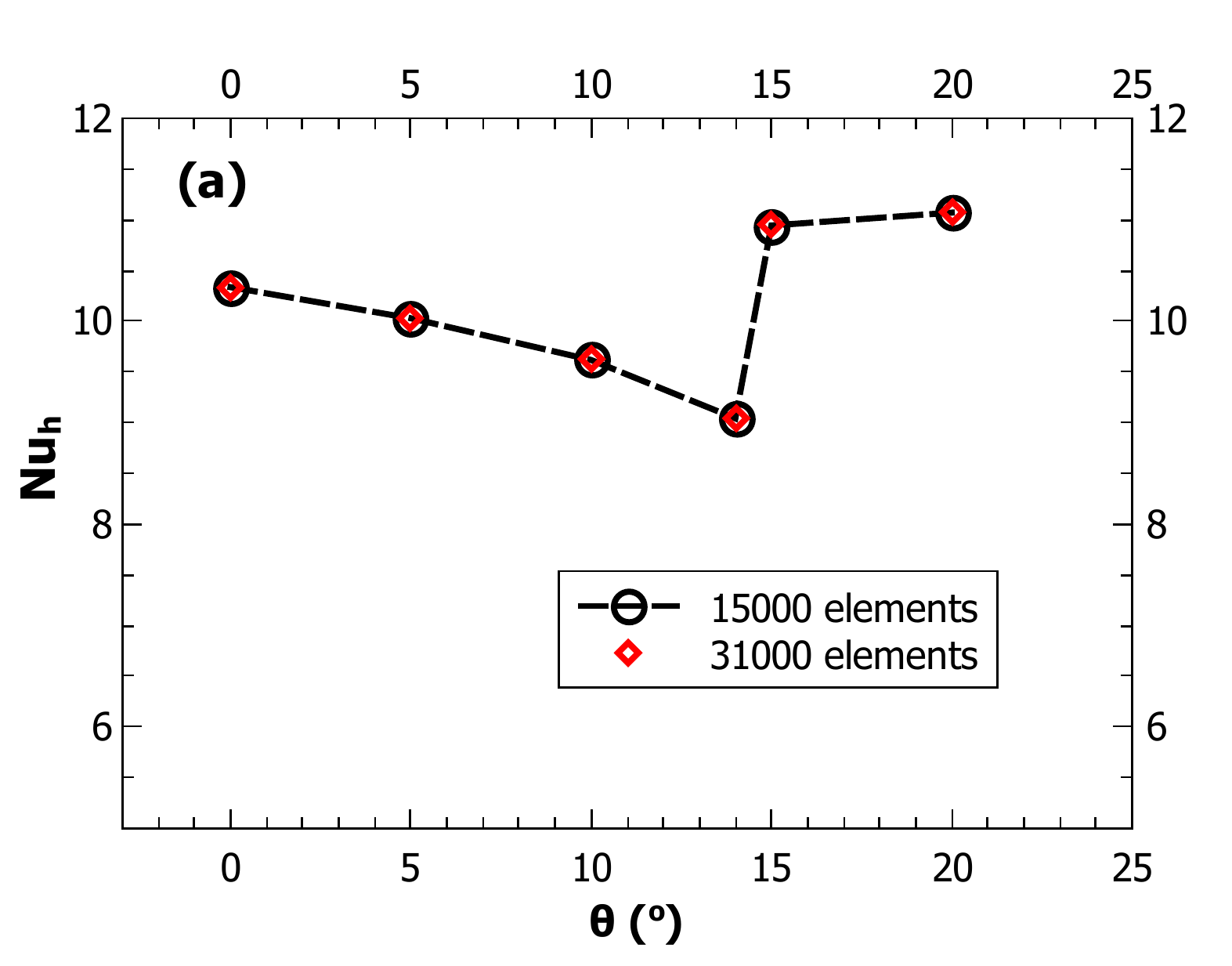}
	\end{subfigure}
	\hspace{\fill}
	\begin{subfigure}[b]{0.49\textwidth}
		\includegraphics[width=1\linewidth]{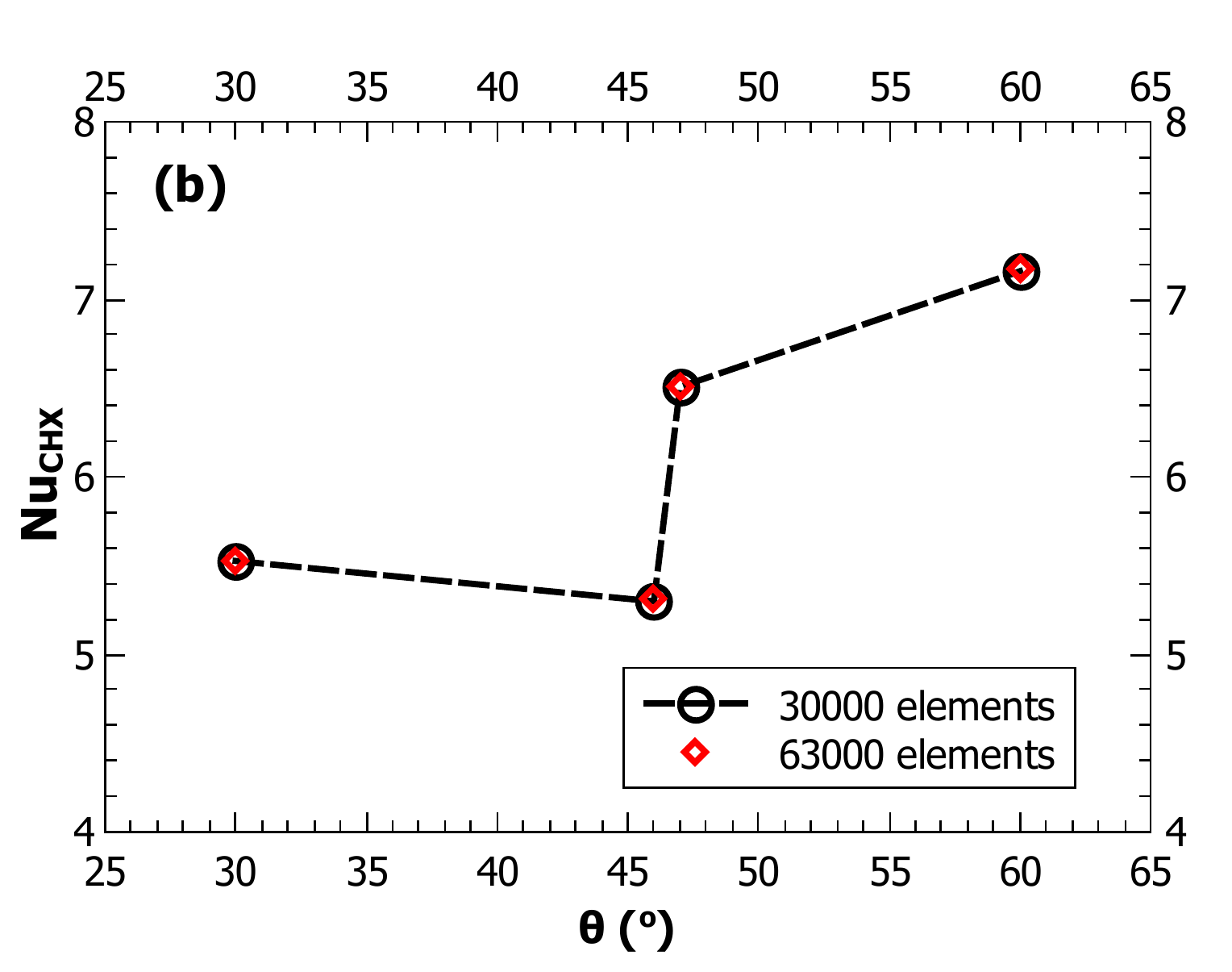}
	\end{subfigure}

	\caption{Mesh independence test : (a) NCL system, (b) CNCL system considering $\theta$ variation from $0^\circ$ to $90^\circ$.  }
	\label{Mesh independence test}
\end{figure}

To ensure that the steady state CFD predictions are free from numerical errors a mesh independence test is conducted. For the mesh independence test $Nu_h$ at the heated section of the NCL system and $Nu_{CHX}$ at the common heat exchange section are considered as parameters. Figure \ref{Mesh independence test} represents the mesh independence test and demonstrates that about 15,000 elements and 30,000 elements are adequate to capture the physics of NCL and CNCL systems, respectively. It is necessary to point out that the mesh independence is deemed to be sucessfully completed only after the point of heat transfer coefficient jump and the magnitude of heat transfer coefficient are completely captured by the refined mesh.

\subsection{Validation of CFD methodology}

\begin{figure}[!htb]
	\centering
	\begin{subfigure}[b]{0.49\textwidth}
		\includegraphics[width=1\linewidth]{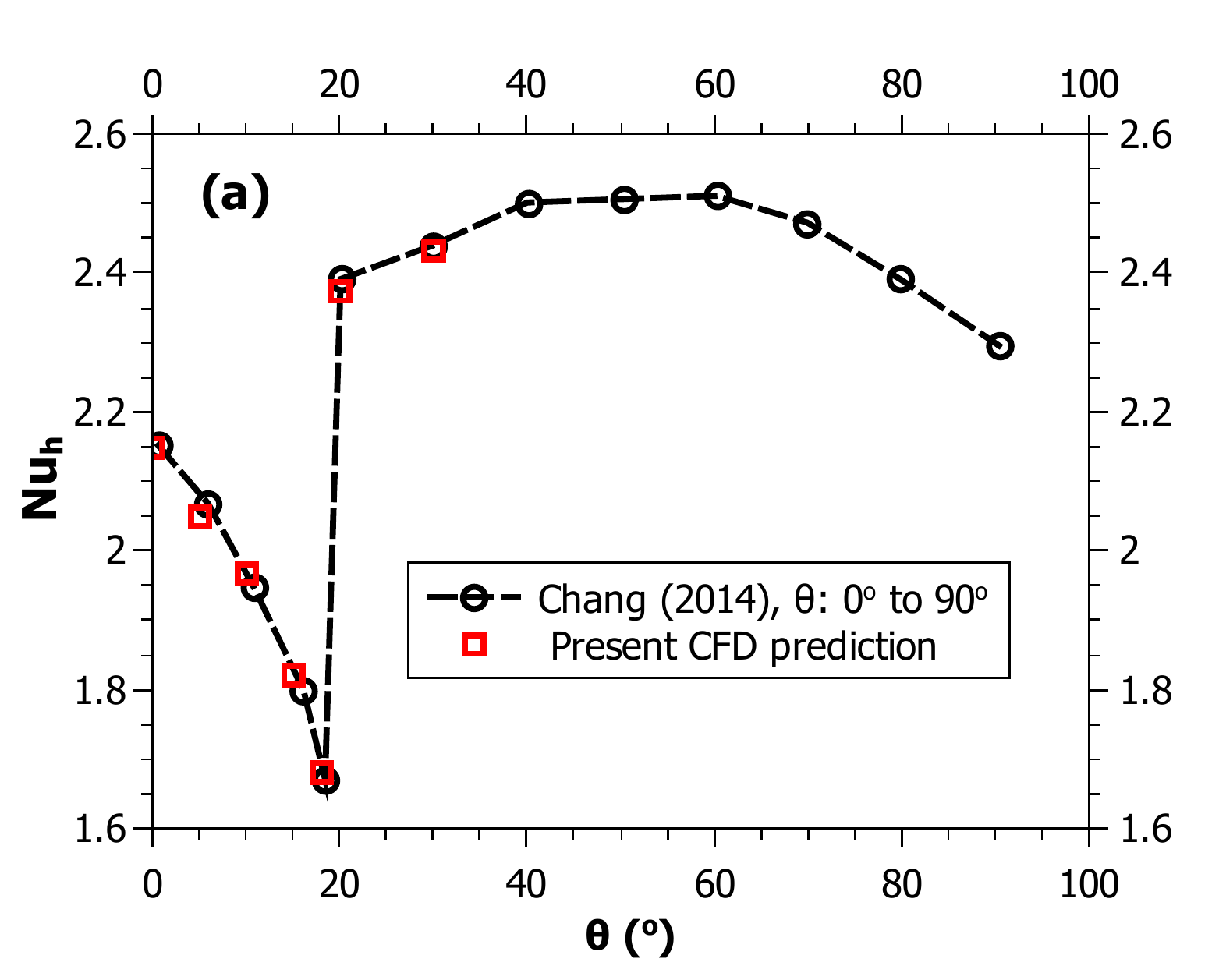}
	\end{subfigure}
	\hspace{\fill}
	\begin{subfigure}[b]{0.49\textwidth}
		\includegraphics[width=1\linewidth]{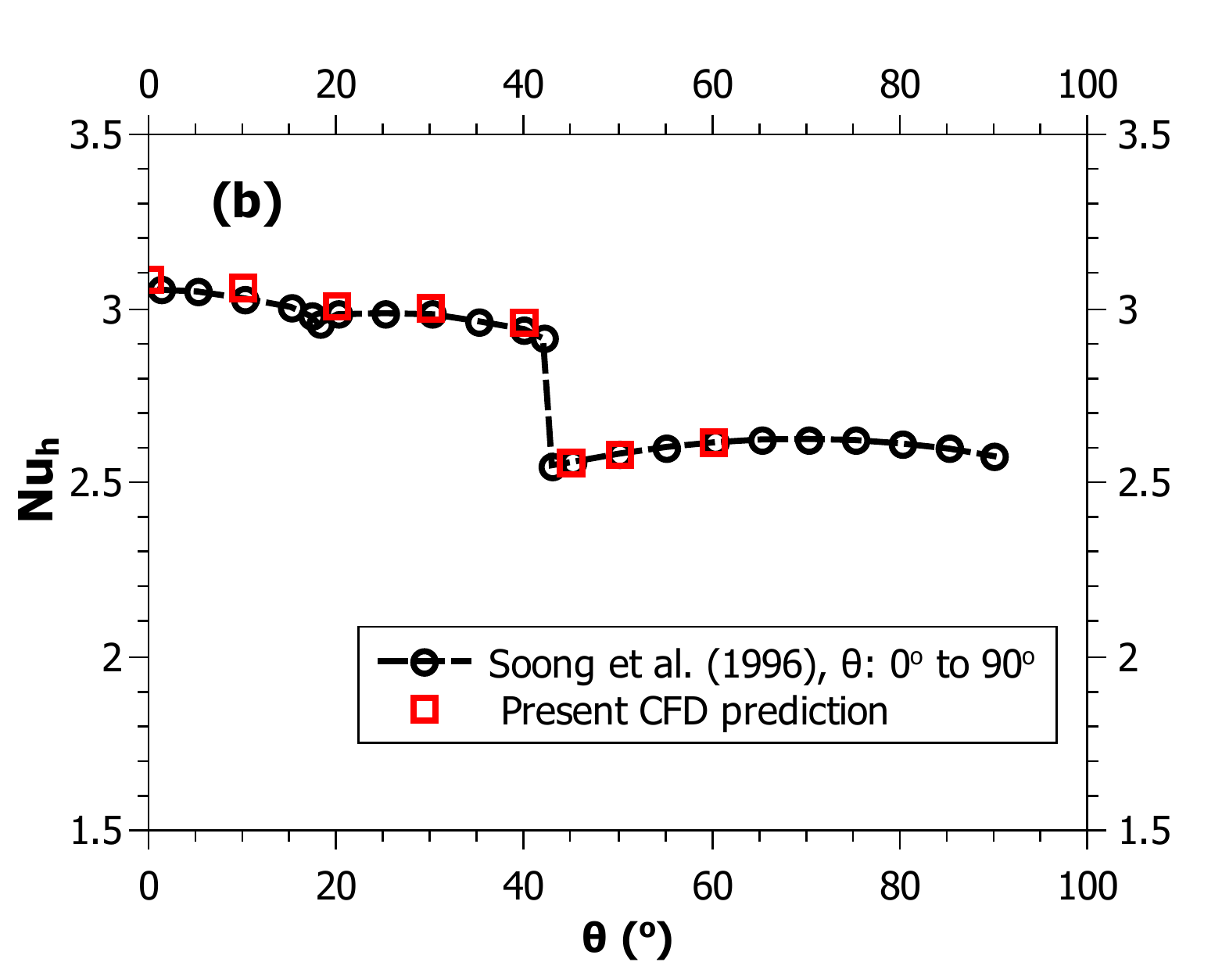}
	\end{subfigure}
		\hspace{\fill}
	\begin{subfigure}[b]{0.49\textwidth}
		\includegraphics[width=1\linewidth]{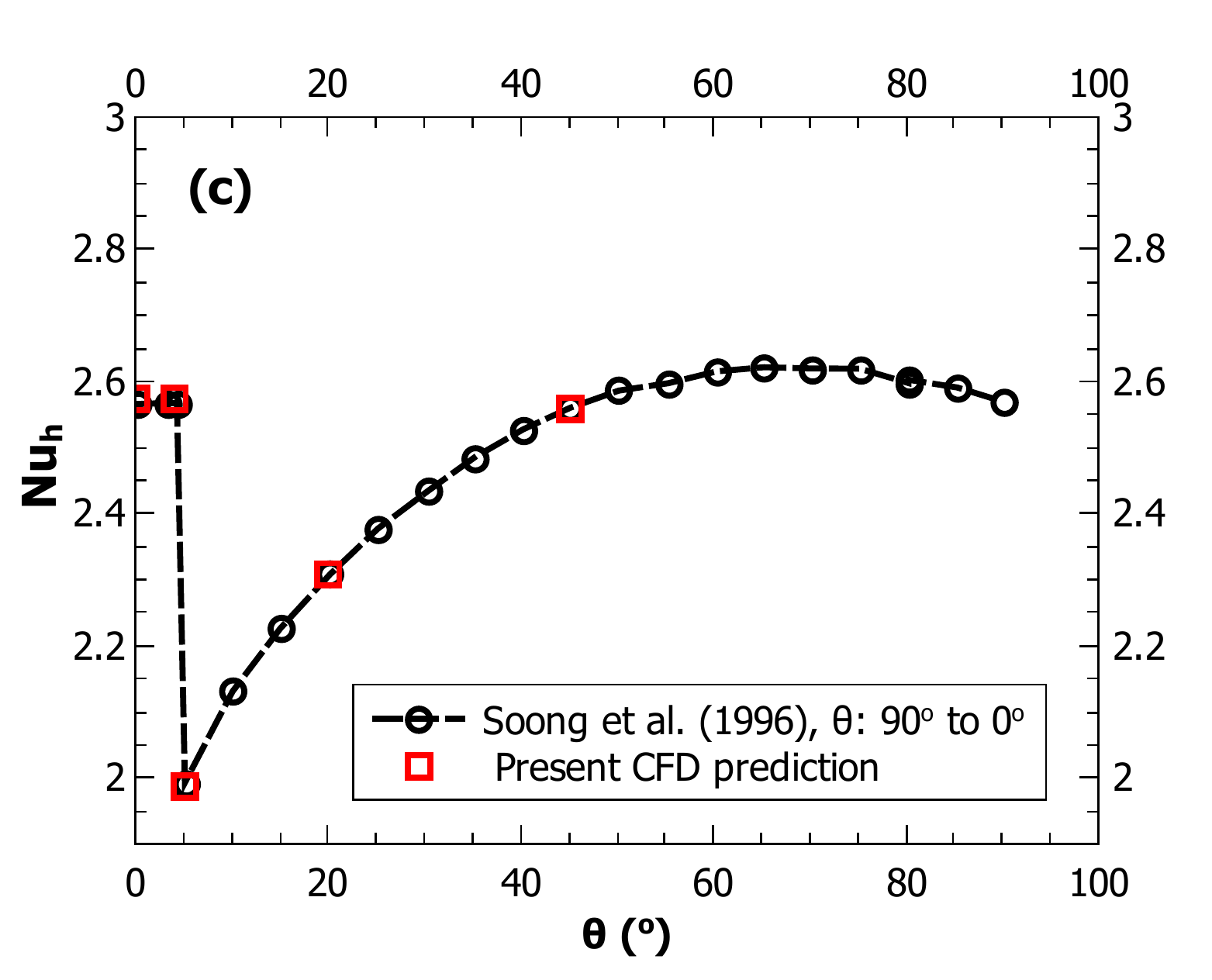}
	\end{subfigure}
	
	\caption{Validation of 2-D CFD with numerical data: (a) Validation with 2-D Cavity study of Chang (2014) \cite{Chang2014} for a cavity of aspect ratio $1$ for $\theta: 0^\circ \; to \; 90^\circ $ and $Ra=10,000$, (b) Validation with 2-D Cavity study of Soong at al. (1996) \cite{Soong1996} for a cavity of aspect ratio $4$ for $\theta: 0^\circ \; to \; 90^\circ $ and $Ra=20,000$, (c) Validation with 2-D Cavity study of Soong at al. (1996) \cite{Soong1996} for a cavity of aspect ratio $4$ for $\theta: 90^\circ \; to \; 0^\circ $ and $Ra=20,000$.}
	\label{Numerical validation}
\end{figure}

To verify the accuracy of the present CFD study to capture the physics of the system, a thorough validation with previous literature is conducted. For the present validation, the thermally excited cavity is considered. For a thorough validation, two different aspect ratios ($AR$) are used, $AR=1$ from Chang (2014) \cite{Chang2014} and $AR=4$ from Soong et al. (1996) \cite{Soong1996}. To demonstrate the ability of the present work to capture the hysteresis effect, a validation is also done with cavity of $AR=4$ for decreasing $\theta$ from $90^\circ$ to $0^\circ$. From Figure \ref{Numerical validation} we observe a good match between literature data and the present CFD prediction confirming the accuracy of the current CFD study in the prediction of dynamics of the buoyancy driven systems.

\section{Results and Discussion}

\subsection{Mechanism of heat transfer coefficient jump in a tilted cavity}

To understand the mechanism by which the heat transfer coefficient jump occurs in a cavity is important as it helps in understanding the process through which the heat transfer coefficient jump occurs in other buoyancy driven systems. This is because of the parallels the cavity system bears with other buoyancy driven systems as explained in the Introduction section. To understand the mechanism involved we study cavity systems of two different aspect ratios, $AR=1$ and $4$. 

\subsubsection{Tilted cavity of $AR=1$}

\begin{figure}[!htb]
	\centering
	\includegraphics[width=0.7\linewidth]{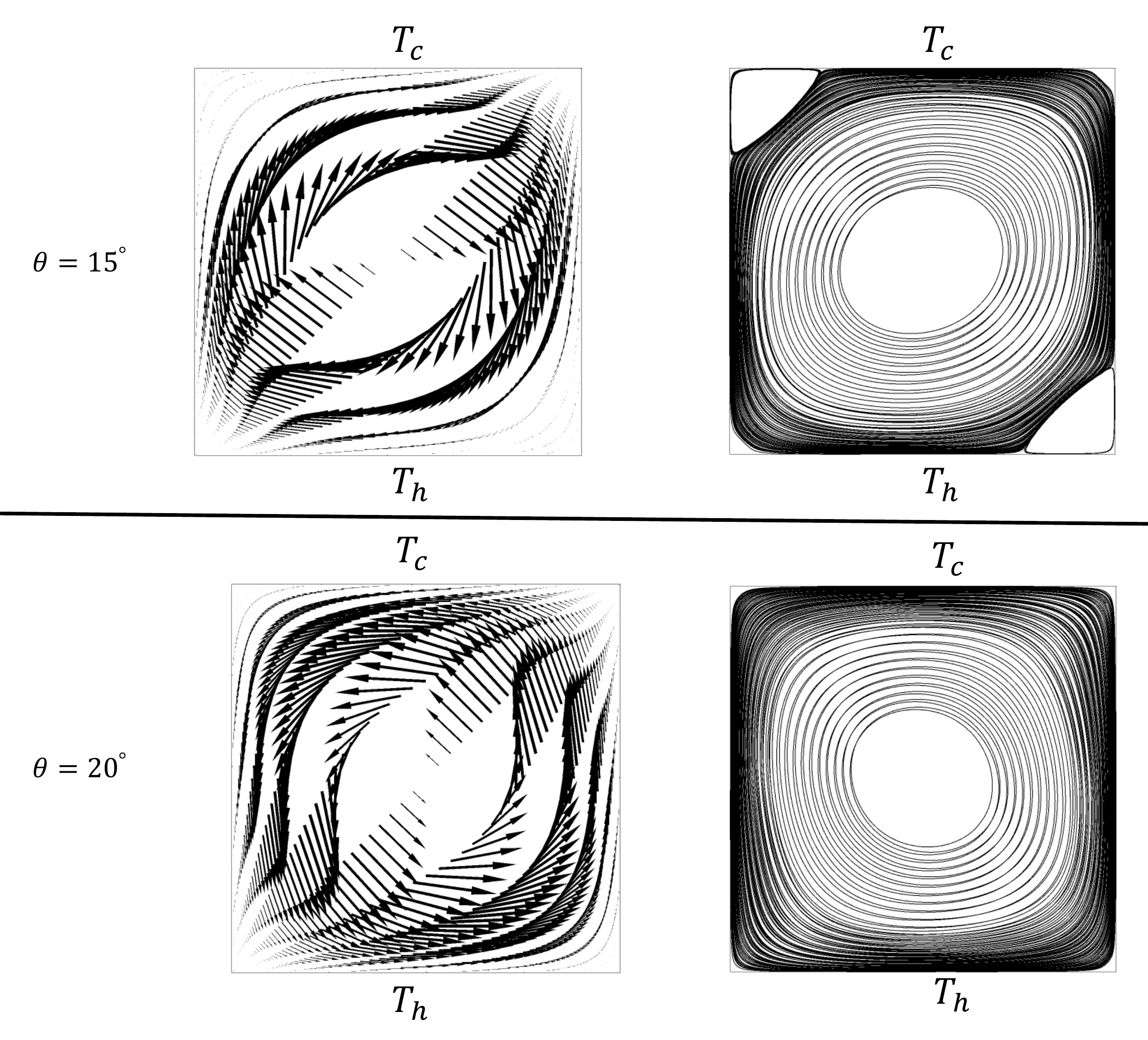}
	\caption{Velocity vectors and streamlines for a cavity of $AR=1$ for $Ra=10,000$ at $\theta=15^\circ$ and $\theta=20^\circ$ considering $\theta$ variation from $0^\circ$ to $90^\circ$.}
	\label{fig:changvelocityvectorandstreamlines}
\end{figure}

From Figure \ref{Numerical validation}(a) we observe that for a cavity of $AR=1$ at $Ra=10,000$ the jump in the heat transfer coefficient occurs at $\theta=18^\circ$, thus if we observe the velocity vectors and streamlines before and after the point of the heat transfer coefficient jump we can understand the phenomena responsible for the heat transfer coefficient jump. From Figure \ref{fig:changvelocityvectorandstreamlines} we observe that there is a difference in both the velocity vectors and the streamline contours. From the velocity vector plot it is evident that a flow direction reversal has occured and from the streamline contour we observe that the smaller vortices have collapsed as we move from $15^\circ$ to $20^\circ$. Thus, we can conclude that the heat transfer coefficient jump occurs by a combination of flow direction reversal and transition from multicellular to unicellular flow patterns for cavities of $AR=1$.

\subsubsection{Tilted cavity of $AR=4$}  

\begin{figure}[!htb]
	\centering
	\includegraphics[width=0.9\linewidth]{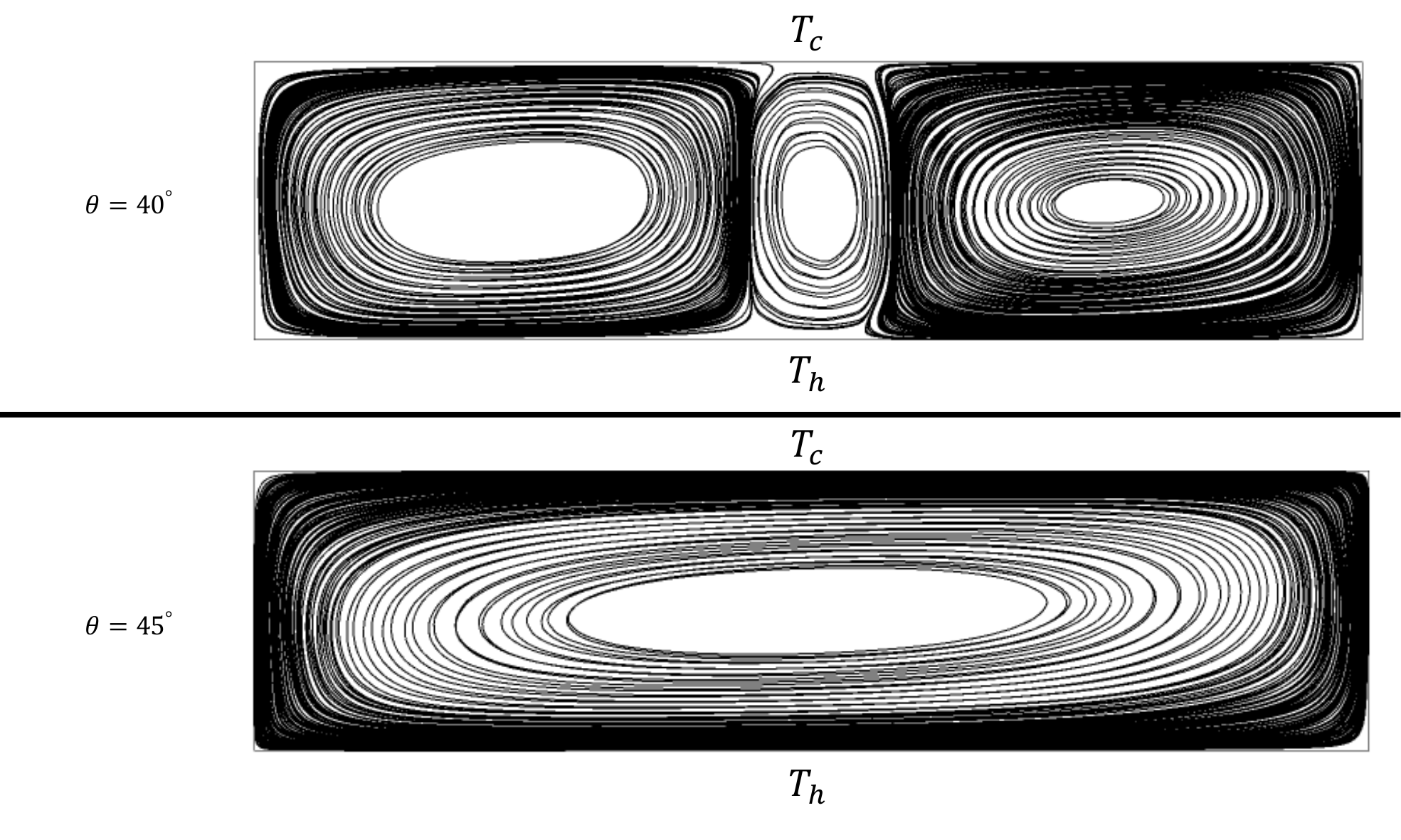}
	\caption{Streamlines for a cavity of $AR=4$ for $Ra=20,000$ at $\theta=40^\circ$ and $\theta=45^\circ$ considering $\theta$ variation from $0^\circ$ to $90^\circ$.}
	\label{fig:SoongStreamlines}
\end{figure}

From Figure \ref{Numerical validation}(b) we observe that for a cavity of $AR=4$ at $Ra=20,000$ the jump in the heat transfer coefficient occurs at $\theta=42^\circ$ . From Figure \ref{fig:SoongStreamlines} we observe that the streamline contours adjacent to the point of jump shift from multicellular to unicellular flow patterns with increasing $\theta$.
This explains the mechanism of the heat transfer coefficient jump in a cavity of $AR=4$. 

The mechanism by which a heat transfer coefficient jump occurs in a cavity was first explained by Soong et al. (1996) \cite{Soong1996} through multicellular to unicellular flow mode transition. This is the most general way to explain a heat transfer coefficient jump or discontinuity in a cavity system, but as we observe for a cavity of $AR=1$ the point at which the jump or discontinuity occurs is accompanied by a net flow direction reversal and a transition from multicellular to unicellular flow. As will be discussed in the upcoming sections, the flow direction reversal in the natural circulation systems such as NCL and CNCL is a prime indicator of the heat transfer coefficient jump.

\subsubsection{Comparison of heat transfer for cavities of $AR=1$ and $AR=4$}

\begin{figure}[!htb]
	\centering
	\begin{subfigure}[b]{0.49\textwidth}
		\includegraphics[width=1\linewidth]{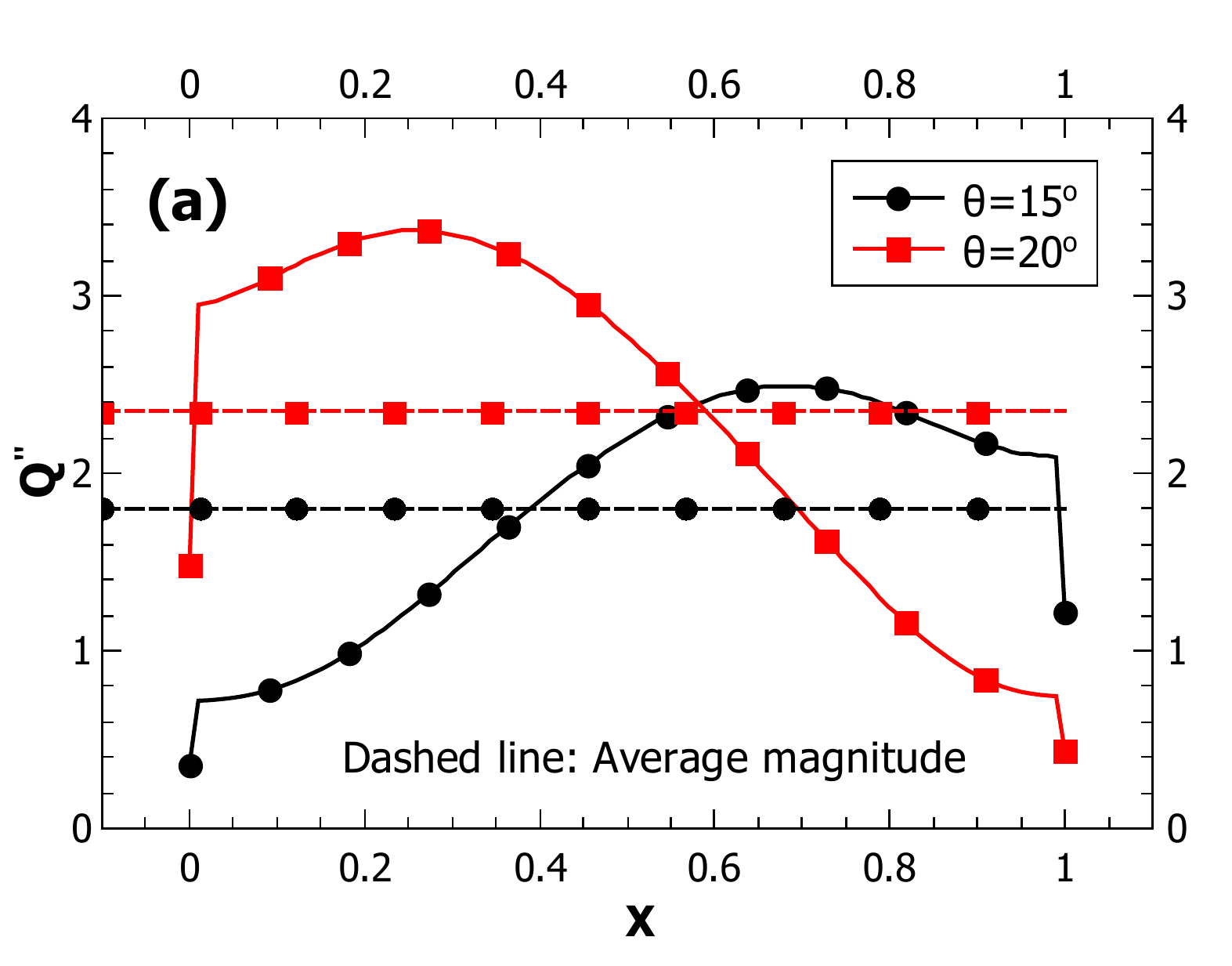}
	\end{subfigure}
	\hspace{\fill}
	\begin{subfigure}[b]{0.49\textwidth}
		\includegraphics[width=1\linewidth]{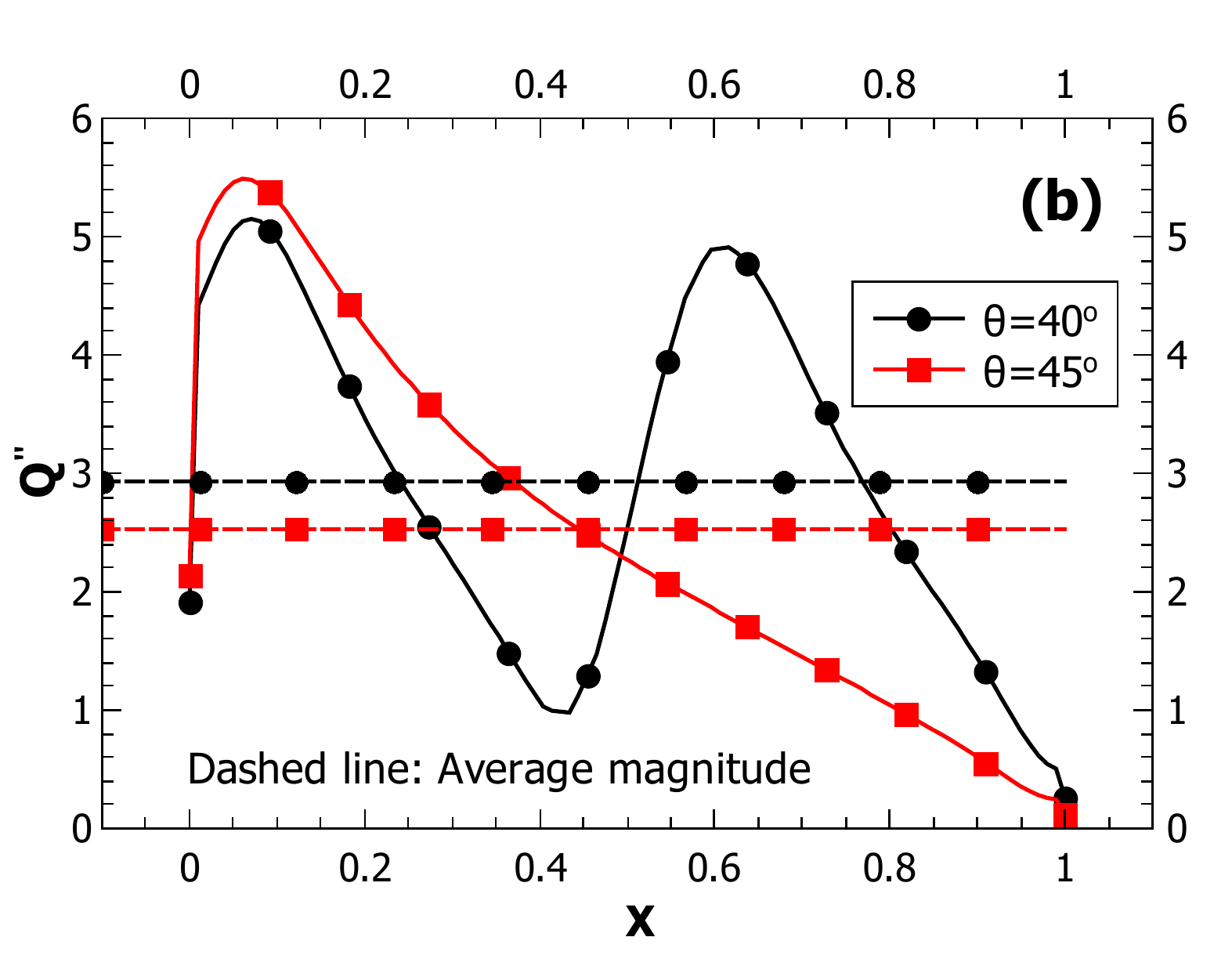}
	\end{subfigure}

	\caption{Comparison of non-dimensional heat flux variation versus non-dimensional length at the heated section of cavity for (a) $AR=1$ and $Ra=10,000$, (b) $AR=4$, $Ra=20,000$. The dashed lines indicate the average magnitude of the non-dimensional heat flux across the non-dimensional length corresponding to the respective angles as indicated by the markers. }
	\label{HeatfluxVsX_Cavity}
\end{figure}

From Figures 6 and 7 it can be observed that at the point of heat transfer coefficient jump there is a flow pattern transition from multicellular to unicellular. While in the case of the cavity system with $AR=1$, this jump in the heat transfer coefficient corresponds to an increase in the Nusselt number (from Figure 5(a)) and for a cavity of $AR=4$, this corresponds to a decrease in the Nusselt number (from Figure 5(b)). This observation implies that the existence of flow vortices does not necessarily increase the magnitude of heat transfer. It may be noted that there is a primary differnce in the nature of vortices observed in both cases. While the vortices in the case of a cavity system with $AR=4$ span the entire height of the cavity and directly transfer the heat from the heated section to the cooled section, the vortices observed in the cavity of $AR=1$ are small and do not contribute significantly to the heat transfer between the hot and cold walls. This conclusion can further be justified from the variation of the local non-dimensional heat transfer coefficient at the heated section as observed in Figure 8.

From Figure \ref{HeatfluxVsX_Cavity}(a) we observe that for $\theta=15^\circ$ the heat transfer decreases along the $x$ direction while the opposite happens for the case of $\theta=20^\circ$. This is primarily because of the flow direction reversal which happens after the heat transfer coefficient jump. At $\theta=15^\circ$ the flow direction is clockwise which implies that the fluid flow at the heated section is in the negative $x$ direction, which means that the fluid which is cooled by the wall approaches the heated wall from $x=1$  to $x=0$ and gets heated up as it progresses in the negative $x$ direction. The decrease in the temperature difference between the fluid and the heated wall along the negative $x$ direction implies a decrease in the heat flux magnitude along the same direction as observed in Figure \ref{HeatfluxVsX_Cavity}(a). Similarly, the non-dimensional heat flux plot shown in Figure \ref{HeatfluxVsX_Cavity}(a) for $\theta=20^\circ$ can be explained in a similar fashion considering the flow direction reversal. It is also to be noted that the flow pattern re-arrangement as observed in Figure \ref{fig:changvelocityvectorandstreamlines} does not have much effect on the non-dimensional heat flux profile along the $x$ direction of the heated section corresponding to the points before and after the heat transfer coefficient jump as represented in Figure \ref{HeatfluxVsX_Cavity}(a). 

Also comparing the non-dimensional heat flux plots shown in Figures \ref{HeatfluxVsX_Cavity}(a) and \ref{HeatfluxVsX_Cavity}(b) we observe a drastic difference in the plots corresponding to the $Q^{\prime \prime}$ distribution at the heated wall corresponding to the points before and after the jump. This allows us to conclusively state that for the considered case of a cavity system with $AR=1$, the flow direction reversal has a greater contribution towards the heat transfer coefficient jump than the flow-pattern re-arrangement.

\subsection{Effect of inclination on an NCL system}

\begin{figure}[!htb]
	\centering
	\begin{subfigure}[b]{0.49\textwidth}
		\includegraphics[width=1\linewidth]{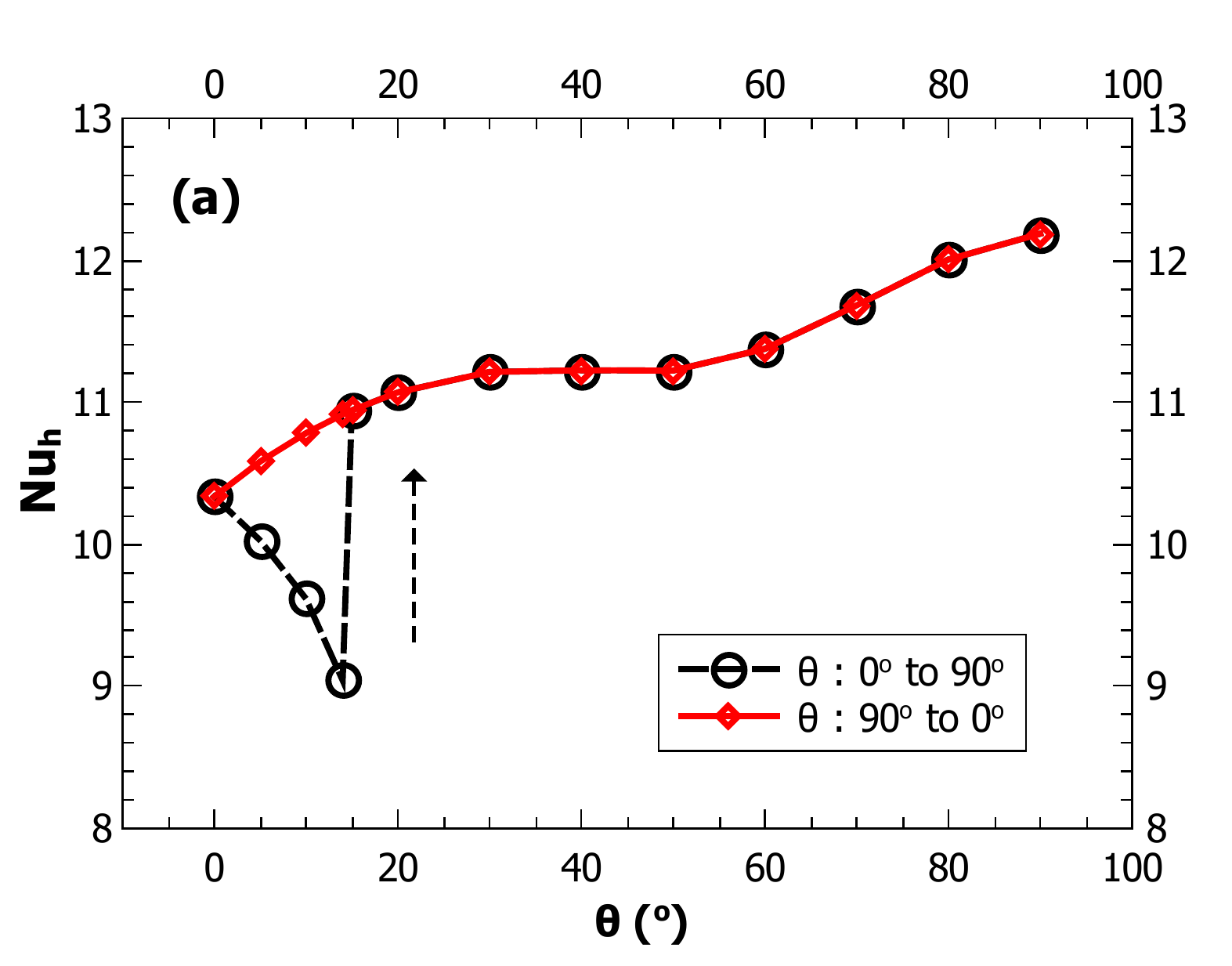}
	\end{subfigure}
	\hspace{\fill}
	\begin{subfigure}[b]{0.49\textwidth}
		\includegraphics[width=1\linewidth]{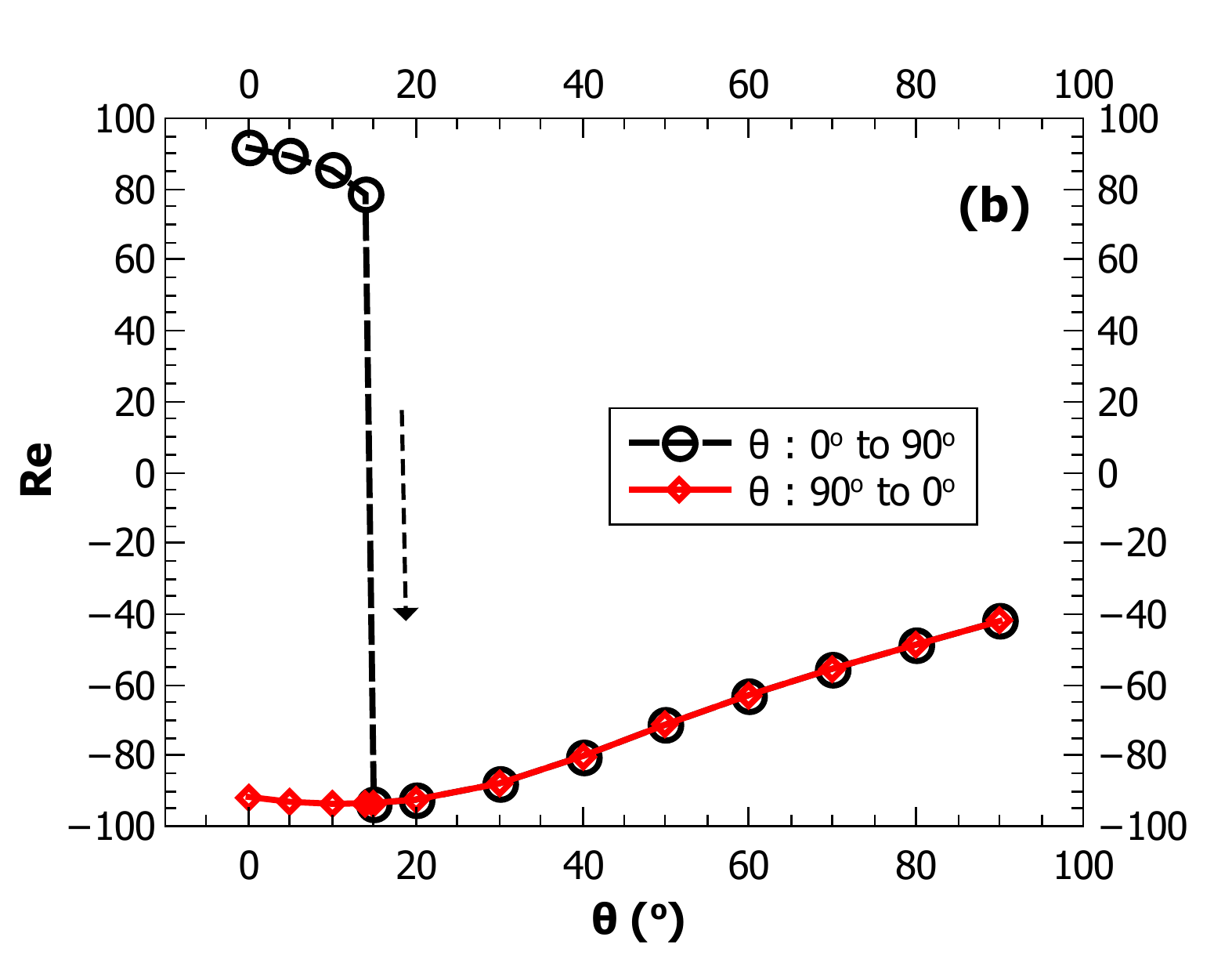}
	\end{subfigure}

	\caption{Effect of inclination on: (a) $Nu_h$, (b) $Re$ of the NCL system for $Ra=1.6 \times 10^5$, $AR=1$, $D/H=0.4$ and $Pr=0.7$. }
	\label{Inclination NCL}
\end{figure}

\begin{figure}[!htb]
	\centering
	\includegraphics[width=0.6\linewidth]{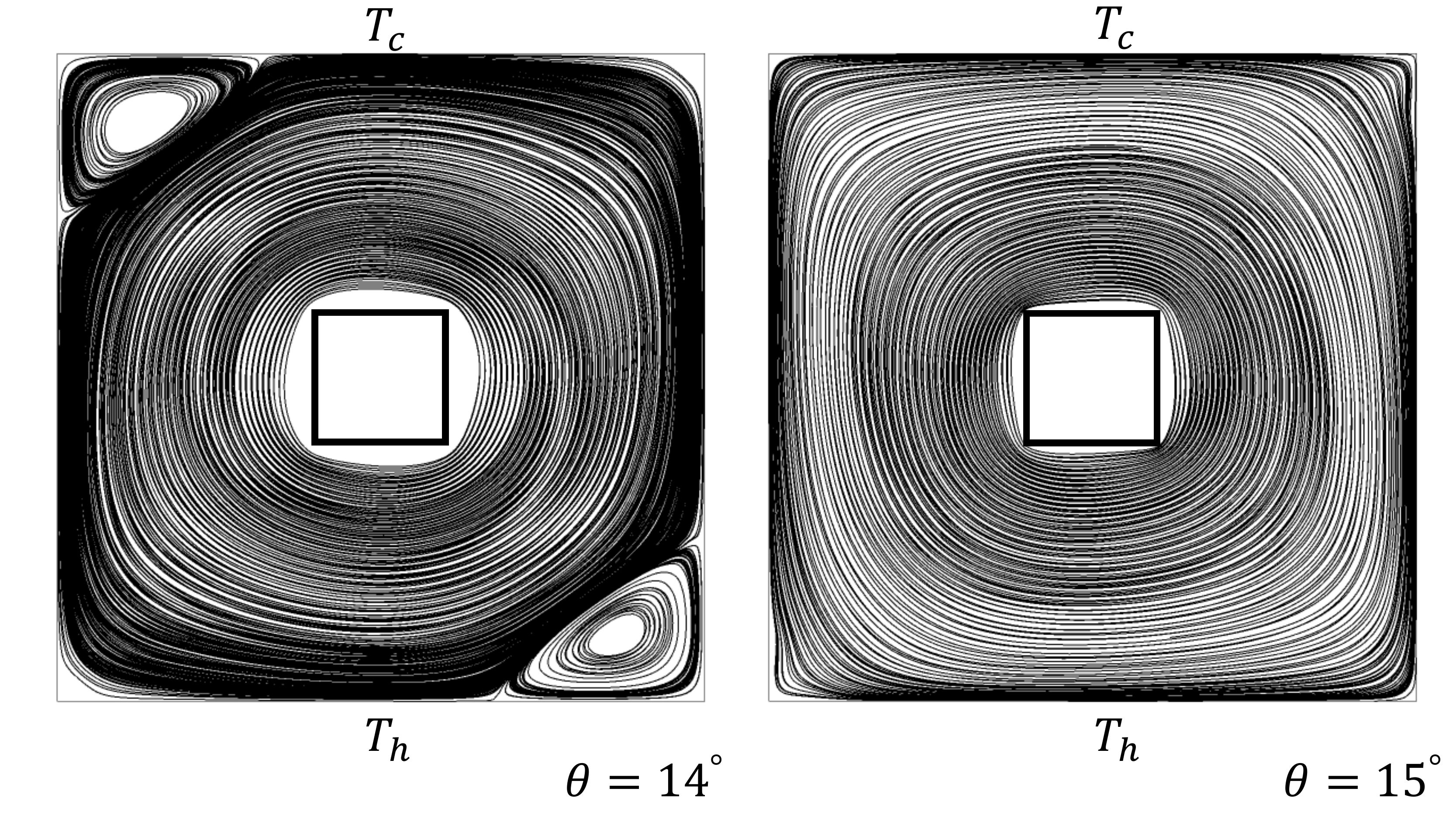}
	\caption{The streamline contours of NCL system at inclinations corresponding to the point before and after the heat transfer coefficient jump for $Ra=1.6 \times 10^5$, $AR=1$, $D/H=0.4$ and $Pr=0.7$.}
	\label{fig:nclstreamlines}
\end{figure}

\begin{figure}[!htb]
	\centering
	\includegraphics[width=0.5\linewidth]{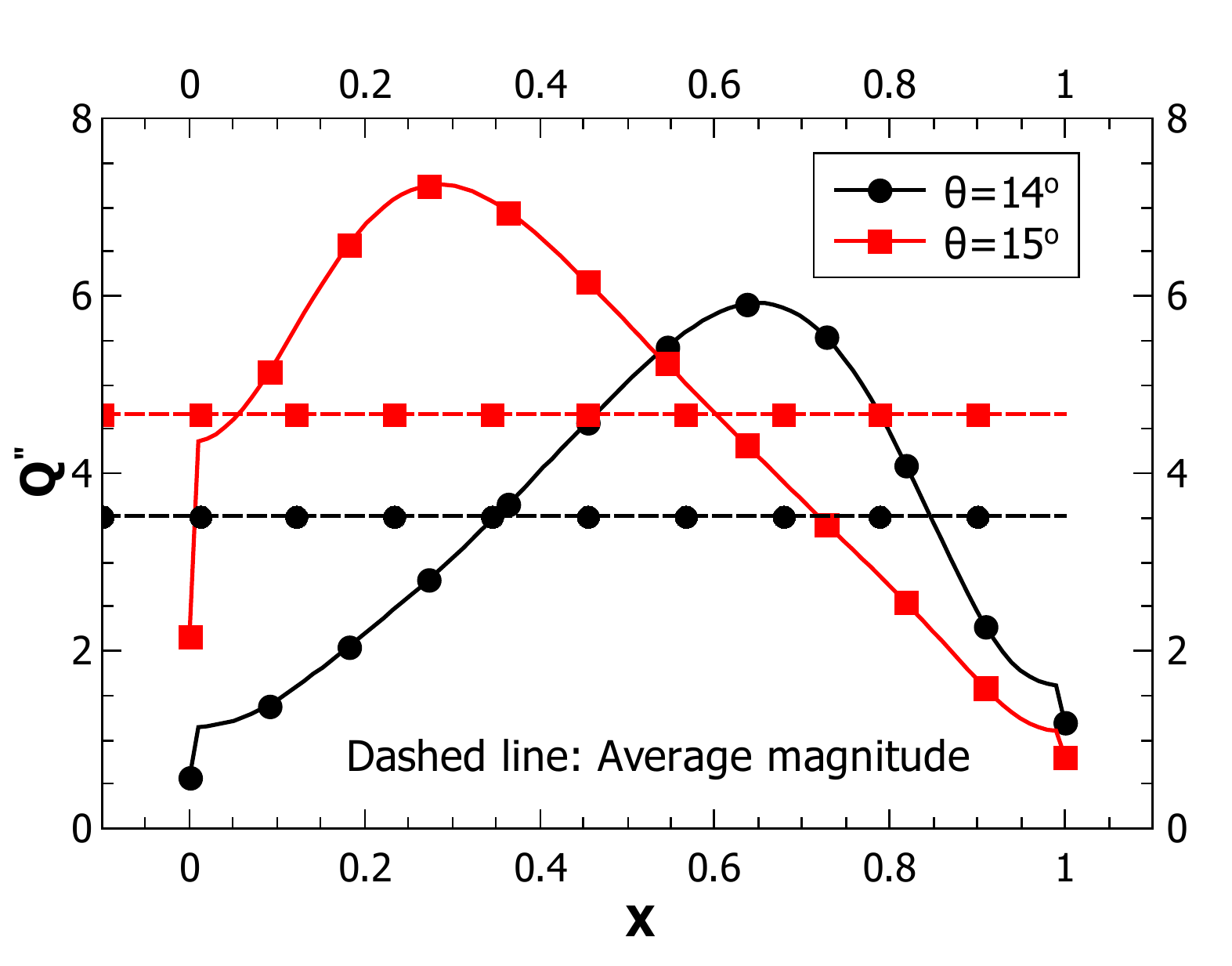}
	\caption{Comparison of non-dimensional heat flux variation versus non-dimensional length at the heated section of NCL for $Ra=1.6 \times 10^5$, $AR=1$, $D/H=0.4$ and $Pr=0.7$ at $\theta=14^\circ$, and $\theta=15^\circ$ . The dashed lines indicate the average magnitude of non-dimensional heat flux across the non-dimensional length corresponding to the respective angles as indicated by the markers.}
	\label{fig:HeatfluxVsX_NCL}
\end{figure}

From Figure \ref{Inclination NCL} we clearly observe that there is a heat transfer coefficient jump in the NCL system. We also clearly observe the hysteresis effect as reflected by the deviation in the data corresponding to the increasing or the decreasing direction of $\theta$. The point at which the heat transfer coefficient jump occurs tallies with the point at which the flow direction reversal occurs. Thus, the heat transfer coefficient jump in the NCL system occurs due to the flow direction reversal. 

The sign associated with Reynolds number ($Re$) in the present study is only to represent the direction of the flow. The clockwise direction is denoted by the positive sign and the anticlockwise direction is denoted by the negative sign. The directions of the flow are assigned based on the direction of the vortex which overlaps with the NCL or CNCL geometry. This notation is consistently used throughout the rest of this work.

The existence of multiple steady states, corresponding to the clockwise and the anti-clockwise directions of flow for $\theta \in (0^\circ,14^\circ)$ is also clearly evident from Figure \ref{Inclination NCL}. The lack of hysteresis phenomena in the decreasing direction of $\theta$ is clearly observed from Figure \ref{Inclination NCL}. This is primarily because, for the direction of  decreasing $\theta$, the buoyancy forces assist the inertial forces in the anti-clockwise flow direction for all $\theta$ and hence no flow-direction reversal is witnessed. Thus, we can conclude that a heat transfer coefficient jump occurs only when the net buoyancy force acting on the system resists the net inertial forces acting upon the system and the point of heat transfer coefficient jump corresponds to the point where the net buoyancy force is equivalent in magnitude to the magnitude of the net opposing inertial and frictional forces.

Figure \ref{fig:nclstreamlines} represents the streamline contours of the NCL system before and after the heat transfer coefficient jump. We can clearly note the flow pattern re-arrangement that has occured. The flow was multi-cellular before the jump, but became unicellular flow after the jump.
A look at the non-dimensional heat flux distribution at the heated wall (Figure \ref{fig:HeatfluxVsX_NCL}) before and after the point of heat transfer coefficient jump reveals no drastic difference because of the flow-pattern re-arrangement but clearly indicates that a flow direction reversal has occurred after the heat transfer coefficient jump as observed earlier for a cavity system of $AR=1$. We also note that the heat transfer coefficient jump corresponds to an increase in the magnitude of the Nusselt number after the jump for the case of ascending $\theta$, thus again implying the fact that the presence of multi-cellular vortices need not necessarily enhance the heat transfer.

\subsection{Effect of inclination on a CNCL system}

\begin{figure}[!htb]
	\centering
	\begin{subfigure}[b]{0.49\textwidth}
		\includegraphics[width=1\linewidth]{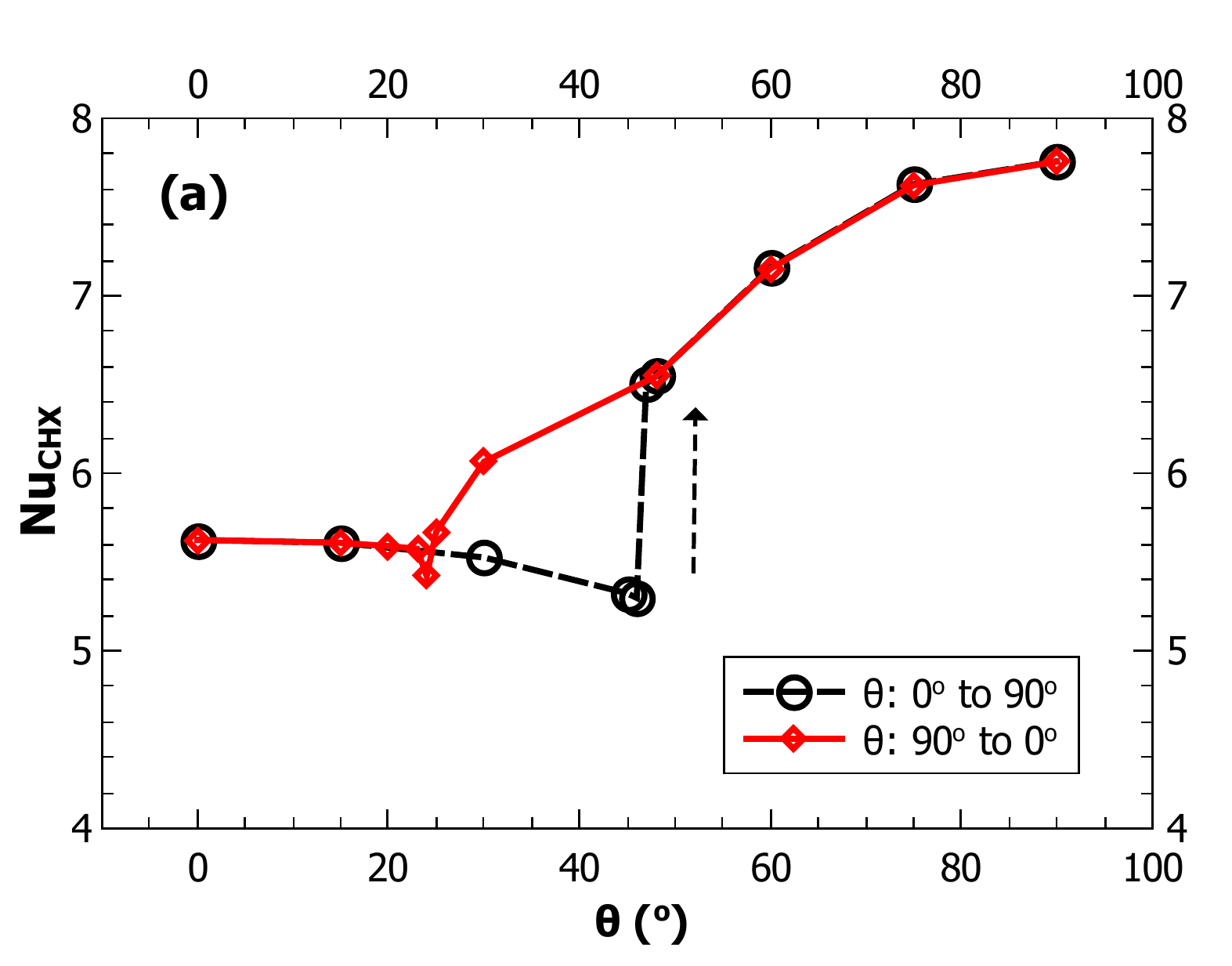}
	\end{subfigure}
	\hspace{\fill}
	\begin{subfigure}[b]{0.49\textwidth}
		\includegraphics[width=1\linewidth]{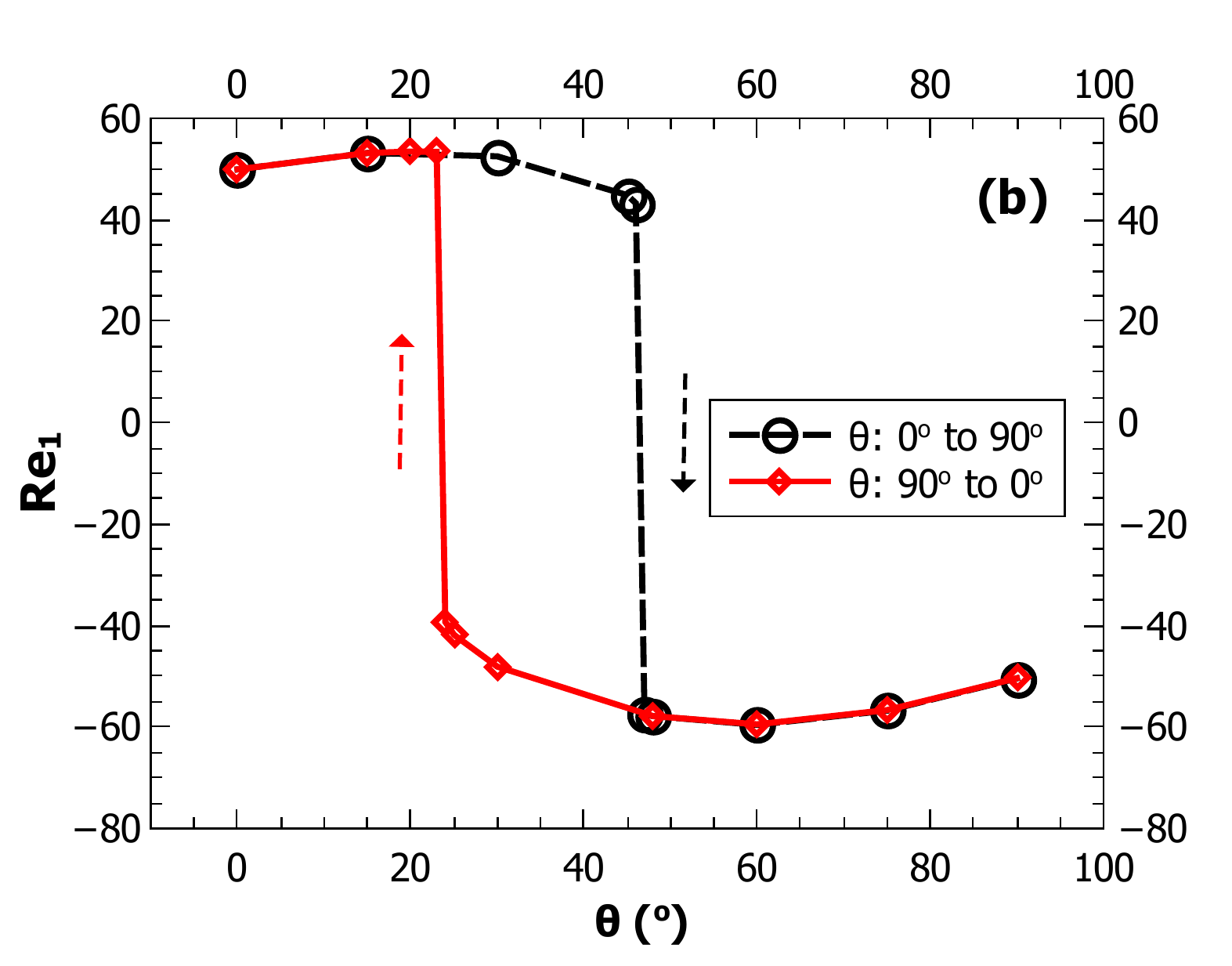}
	\end{subfigure}

	\caption{Effect of inclination on: (a) $Nu_{CHX}$, (b) $Re_1$, of the CNCL system for $Ra=1.6 \times 10^5$, $AR=1$, $D/H=0.4$ and $Pr=0.7$. }
	\label{Inclination CNCL}
\end{figure}

From Figure \ref{Inclination CNCL} we note that a heat transfer coefficient jump occurs at $\theta=46^\circ$ for an increasing $\theta$ and at $\theta=24^\circ$ for a decreasing $\theta$. Once again it can be seen that the heat transfer coefficient jump observed in Figure \ref{Inclination CNCL}(a) is accompanied by a flow direction reversal as observed in Figure \ref{Inclination CNCL}(b). Thus, we can conclude that a heat transfer coefficient jump or discontinuity in a natural circulation system such as an NCL or a CNCL is linked with the flow direction reversal.

\begin{figure}[!htb]
	\centering
	\includegraphics[width=\linewidth]{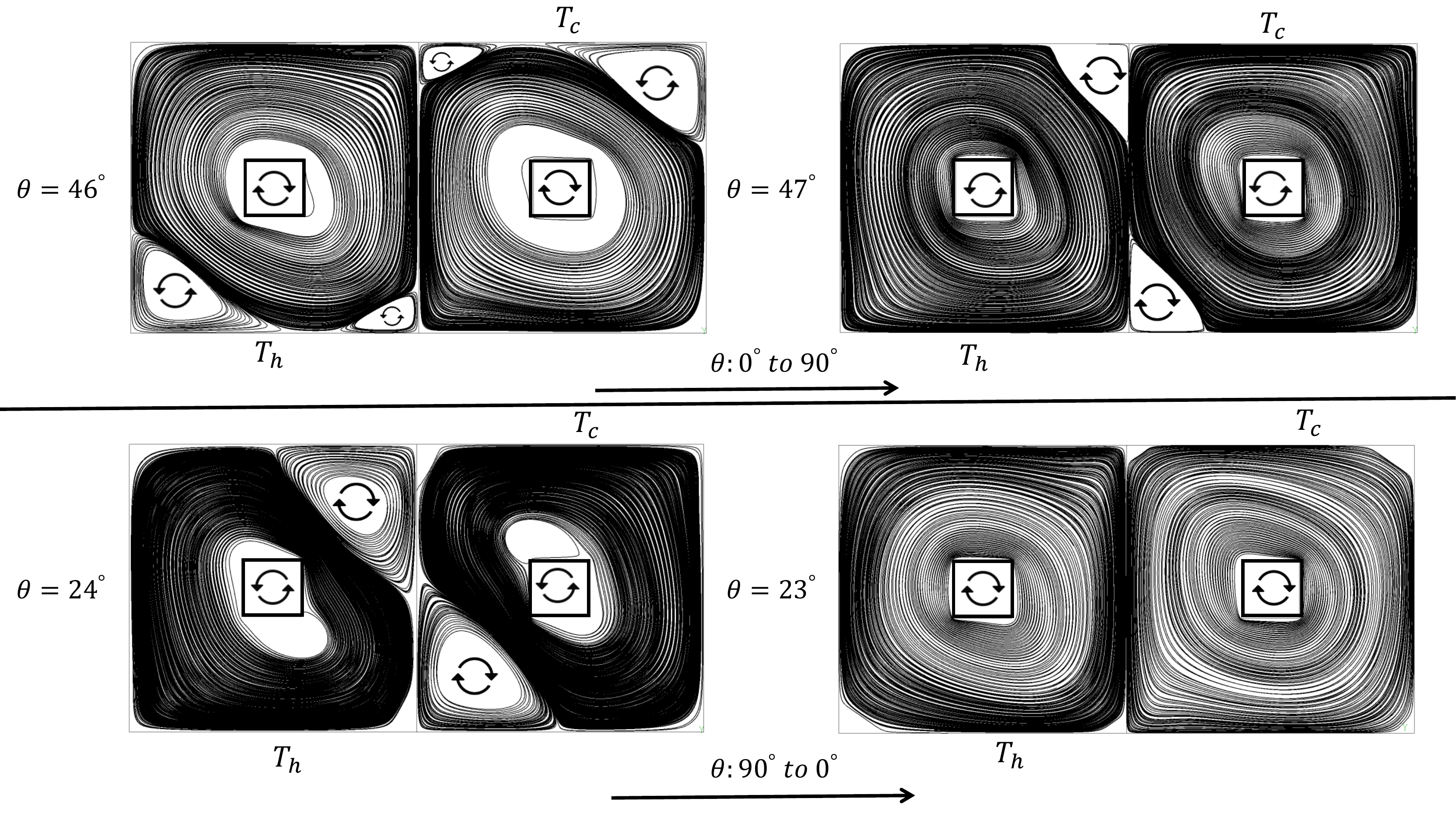}
	\caption{Streamlines of CNCL system for $\theta: 0^\circ\; to\; 90^\circ$ and $\theta: 90^\circ\; to \;0^\circ$ for the cases adjacent to the point of flow direction reversal for $Ra=1.6 \times 10^5$, $AR=1$, $D/H=0.4$ and $Pr=0.7$. The circular arrows at the center of each of the figure denote the direction of the vortex aligned with the geometry of the system, and the remaining circular arrows denote the direction of the smaller vortices.}
	\label{fig:cnclstreamlinesbothdirections}
\end{figure}

There is a difference in the magnitude of heat transfer coefficient jump when approached from different directions. This may be understood by looking at the streamline contours in both directions corresponding to the point of the jump. Figure \ref{fig:cnclstreamlinesbothdirections} shows the streamline contours from which it can be observed that the heat transfer at the coupled wall is affected by the smaller vortices and the larger vortex which is aligned with the system geometry ( this is the vortex which contributes to the circulation within the loop). For the case of increasing $\theta$ we observe that the net configuration at the coupled wall before and after the jump is a combination of both parallel and counterflow configurations. Whereas for the case of decreasing $\theta$ we observe that the configuration is almost completely parallel flow at $\theta=24^\circ$ and the configuration at the coupled wall after the jump or discontinuity i.e. at $\theta=23^\circ$ is completely counterflow. This explains the nature of hysteresis and the magnitude of the jump observed in CNCL systems when the inclination is varied in the direction of increasing and decreasing $\theta$.

\begin{figure}[!htb]
	\centering
	\begin{subfigure}[b]{0.49\textwidth}
		\includegraphics[width=1\linewidth]{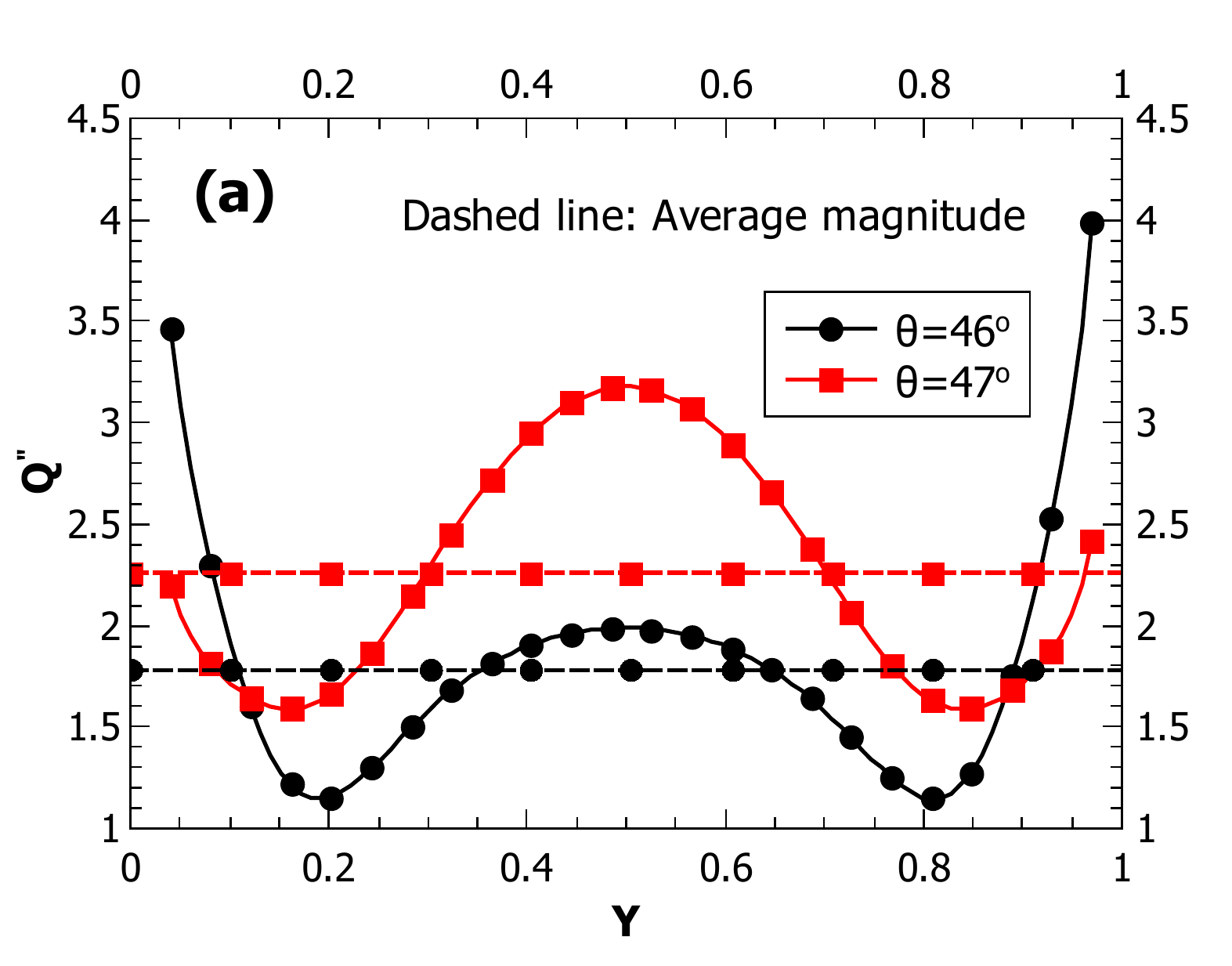}
	\end{subfigure}
	\hspace{\fill}
	\begin{subfigure}[b]{0.49\textwidth}
		\includegraphics[width=1\linewidth]{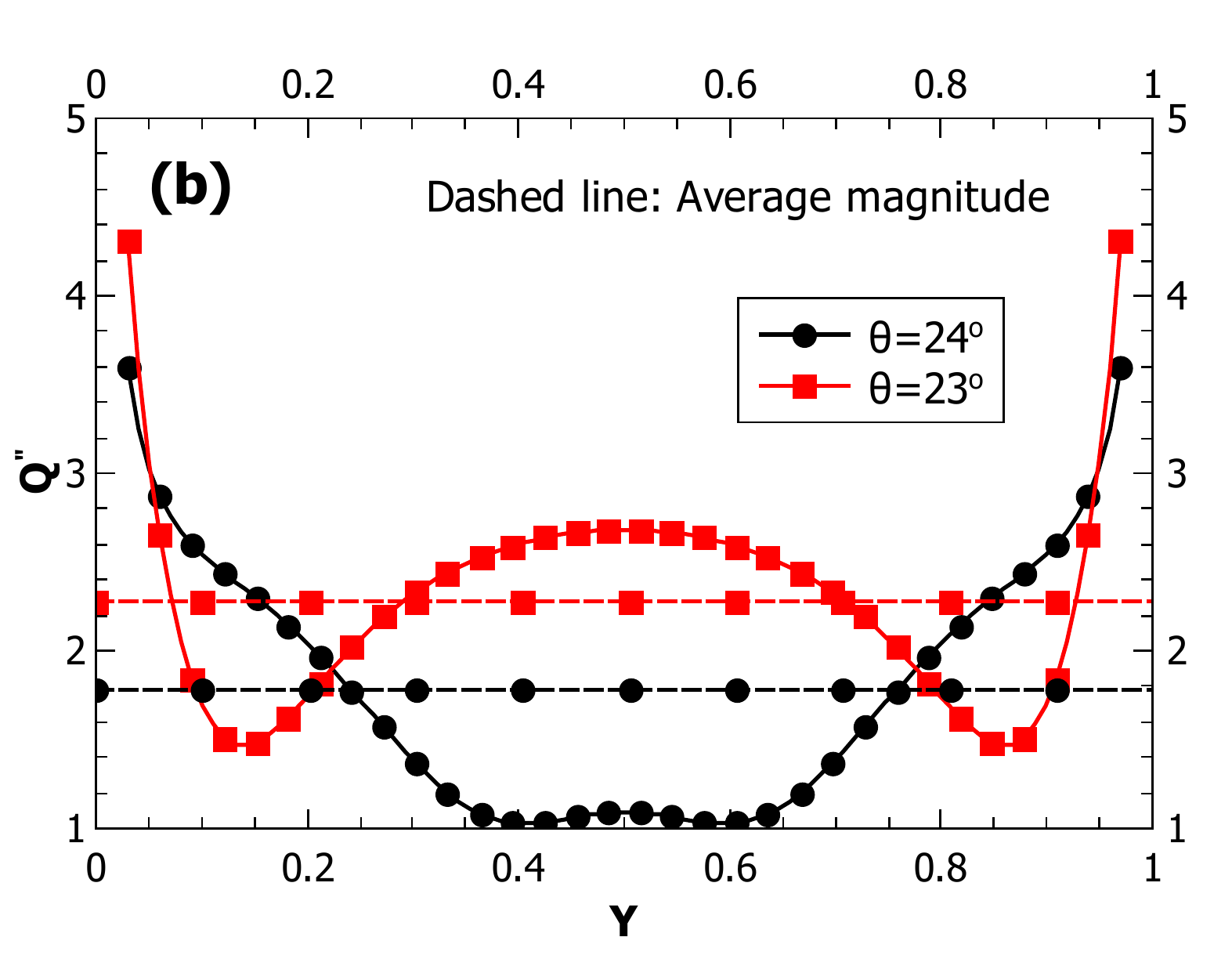}
	\end{subfigure}

	\caption{Local non-dimensional heat flux  at the coupled heat exchange section of CNCL for $Ra=1.6 \times 10^5$, $AR=1$, $D/H=0.4$ and $Pr=0.7$ for the point corresponding to the heat transfer coefficient jump for (a) ascending direction of $\theta$ (b) descending direction of $\theta$. The dashed lines indicate the average magnitude of the non-dimensional heat flux across the non-dimensional length corresponding to the respective angles as indicated by the markers. }
	\label{HeatFluxVsX_CNCL}
\end{figure}

The variation of local non-dimensional heat flux at the coupled heat exchange section of the CNCL shown in Figure \ref{HeatFluxVsX_CNCL} correspond to the respective streamline contours shown in Figure \ref{fig:cnclstreamlinesbothdirections}. It can be observed from Figure \ref{HeatFluxVsX_CNCL}(b)  that the profiles of distribution of $Q^{\prime \prime}$ are different for $\theta=23^\circ$ and $\theta=24^\circ$. This corresponds to the shift from parallel flow configuration to counterflow configuration at the common heat exchange section for the case of descending direction of $\theta$ with the counterflow configuration at the common heat exchange section ($\theta=23^\circ \; for \; \theta:90^\circ \; to \; 0^\circ$) having a greater magnitude.

\subsection{Why was the heat transfer coefficient jump not reported in literature on the tilted NCL systems?}

A thorough review of literature available on the  tilted NCL systems is presented in the Introduction section. From the available literature it is observed that  the heat transfer coefficient jump has not been reported in any of the study so far, to the best of the authors' knowledge. This section investigates and provides the explanation as to why the phenomenon of the heat transfer coefficient jump was not identified in the tilted NCL systems earlier.

\begin{table}[!htb]
	\caption{Parameters considered in the previous studies on the tilted NCL systems.}
	\begin{tabular}{|l|l|l|l|l|l|l|}
		\hline
		\textbf{Sr.no} & \textbf{Author}                                                           & \begin{tabular}[c]{@{}l@{}} \textbf{Minimum}\\ \textbf{inclination}\\ \textbf{step size} \end{tabular} & \begin{tabular}[c]{@{}l@{}}\textbf{Axes of} \\ \textbf{rotation}\end{tabular} & \textbf{Heater-cooler position}   & \begin{tabular}[c]{@{}l@{}} \textbf{Investigation} \\ \textbf{of Hysteresis} \\ \textbf{effect}  \end{tabular}  & \begin{tabular}[c]{@{}l@{}} \textbf{Type of} \\ \textbf{study}  \end{tabular}           \\ \hline
		1 & \begin{tabular}[c]{@{}l@{}}Krishnani \\ and Basu (2017)\end{tabular}    & 15                                                                      & z                                                           & \begin{tabular}[c]{@{}l@{}}Opposite heater-cooler \end{tabular}   & No & Transient                              \\ \hline
		2 & Ramos et al. (1990)                                                     & N.A                                                                     & z                                                           & \begin{tabular}[c]{@{}l@{}}Opposite heater-cooler \end{tabular}    & Yes & Steady state                             \\ \hline
		3 & Acosta et al. (1987)                                                   & 15                                                                      & z                                                           & \begin{tabular}[c]{@{}l@{}}Opposite heater-cooler \end{tabular}    & Yes & Steady state                             \\ \hline
		4 & Misale et al. (2007)                                                   & N.A                                                                     & x                                                           & \begin{tabular}[c]{@{}l@{}}Opposite heater-cooler \end{tabular}  & No & Transient                               \\ \hline
		5 & C. Tian et al. (2017)                                                  & 10                                                                      & z                                                           & \begin{tabular}[c]{@{}l@{}}Adjacent heater-cooler \end{tabular} & No & Steady state \\ \hline
		6 & \begin{tabular}[c]{@{}l@{}}Garibaldi and\\   Misale (2008)\end{tabular} & N.A                                                                     & x                                                           & \begin{tabular}[c]{@{}l@{}}Opposite heater-cooler \end{tabular}  & No & Transient                               \\ \hline
		7 & D.N. Basu et al. (2013)                                                 & 30                                                                      & z                                                           & \begin{tabular}[c]{@{}l@{}}Opposite heater-cooler \end{tabular}    & No & Transient                             \\ \hline
		8 & H. Bouali et al. (2006)                                                  & 15                                                                      & z                                                           & \begin{tabular}[c]{@{}l@{}}Opposite heater-cooler \end{tabular}   & No & Steady state                              \\ \hline
		9 & Misale et al. (2005)                                                    & 30                                                                      & x                                                           & \begin{tabular}[c]{@{}l@{}}Opposite heater-cooler \end{tabular}     & No & Transient                            \\ \hline
		10 & Zhu et al. (2017)                                                      & 10                                                                      & z                                                           & \begin{tabular}[c]{@{}l@{}}Adjacent heater-cooler \end{tabular}  & No & Steady state  \\ \hline
	\end{tabular}
\label{table:Literature}
\end{table}

Table \ref{table:Literature} provides a list of important parameters which were used for the study of inclination effect on NCL systems in the available literature. The possible reasons due to which the heat transfer coefficient jump in the tilted NCL systems may not have been reported are:

\begin{enumerate}
	\item Considering only the cases where the buoyancy forces assist the flow: As mentioned in section.4.2, the flow direction reversal happens only when the net buoyancy forces for the current angle oppose the flow (the steady state solution for the previous angle used as the initial condition for the current angle) within the loop. Thus, if only the cases where buoyancy assists the flow with the initial zero flow field are considered, then no flow direction reversal happens and this consequently results in no observation of the heat transfer coefficient jump in the tilted NCL systems. This may not be intentional but for a transient study which is initiated with the fluid initially at rest within the tilted NCL, the flow direction is dictated by the buoyancy forces and thus the flows observed are naturally always assisted by buoyancy. This is evident from Figure 15(a) and 15(b) for the cases of intial flow field being zero for all angles and the intial flow field being the steady state solution of the previous angle, respectively.
	\item Lack of hysteresis study: From Table
	1, it can be noted that apart from Acosta et al. \cite{Acosta1987} and Ramos et al. \cite{Ramos1990} no other study has considered the hysteresis effect. This may possibly  be due to the focus of the study being the transient analysis. If the hysteresis study were conducted for the steady state cases with the initial condition being the steady state solution of the previous angle, then a case where the buoyancy forces resulting from the current angle resist the initial flow within the loop might have been encountered and the heat transfer coefficient jump might have been observed.
	\item Large step size of $\Delta \theta$ : From Table 1 we note that minimum step size of inclination is more than or equal to $10^\circ$. If a larger step size of $\Delta \theta$ is considered then the resolution of the study may be inadequate to capture the heat transfer coefficient jump. This may be observed from Figure \ref{fig:stepsizeeffectoninclinationnclystem}(b), where if we consider a $\Delta \theta > 10^\circ$ for the considered case, the point at which the heat transfer coefficient jump occurs is not captured. It can be observed that  $\Delta \theta$ will not affect the heat transfer coefficient after the point of jump. For the case of $\Delta \theta=1^\circ, \; 0.5^\circ$, only the points around the zone of the heat transfer coefficient jump are indicated, for clarity, as otherwise it would overlap with the points corresponding to the other step sizes.
	
	\begin{figure}[!htb]
	\centering
	\begin{subfigure}[b]{0.49\textwidth}
		\includegraphics[width=1\linewidth]{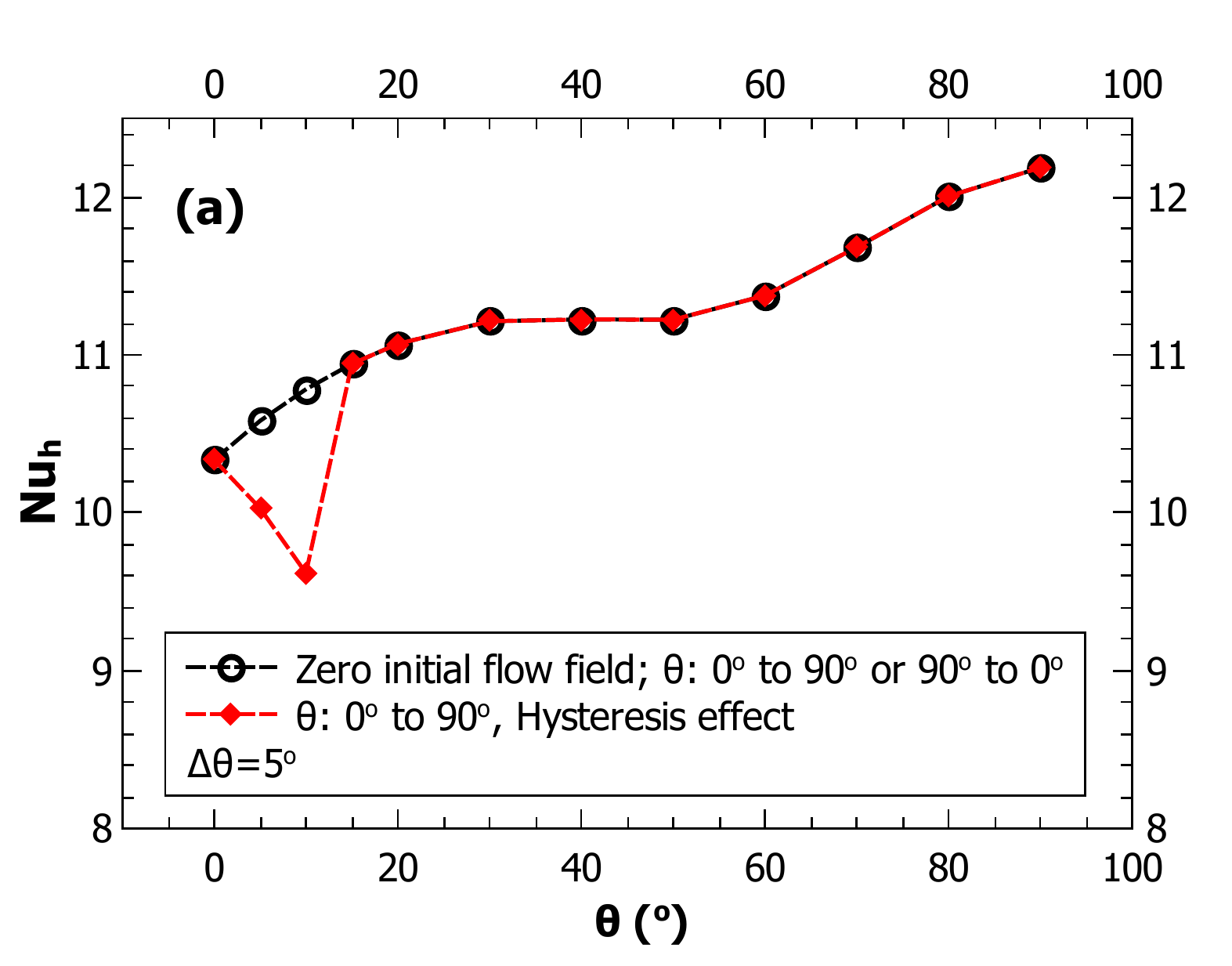}
	\end{subfigure}
	\hspace{\fill}
	\begin{subfigure}[b]{0.49\textwidth}
		\includegraphics[width=1\linewidth]{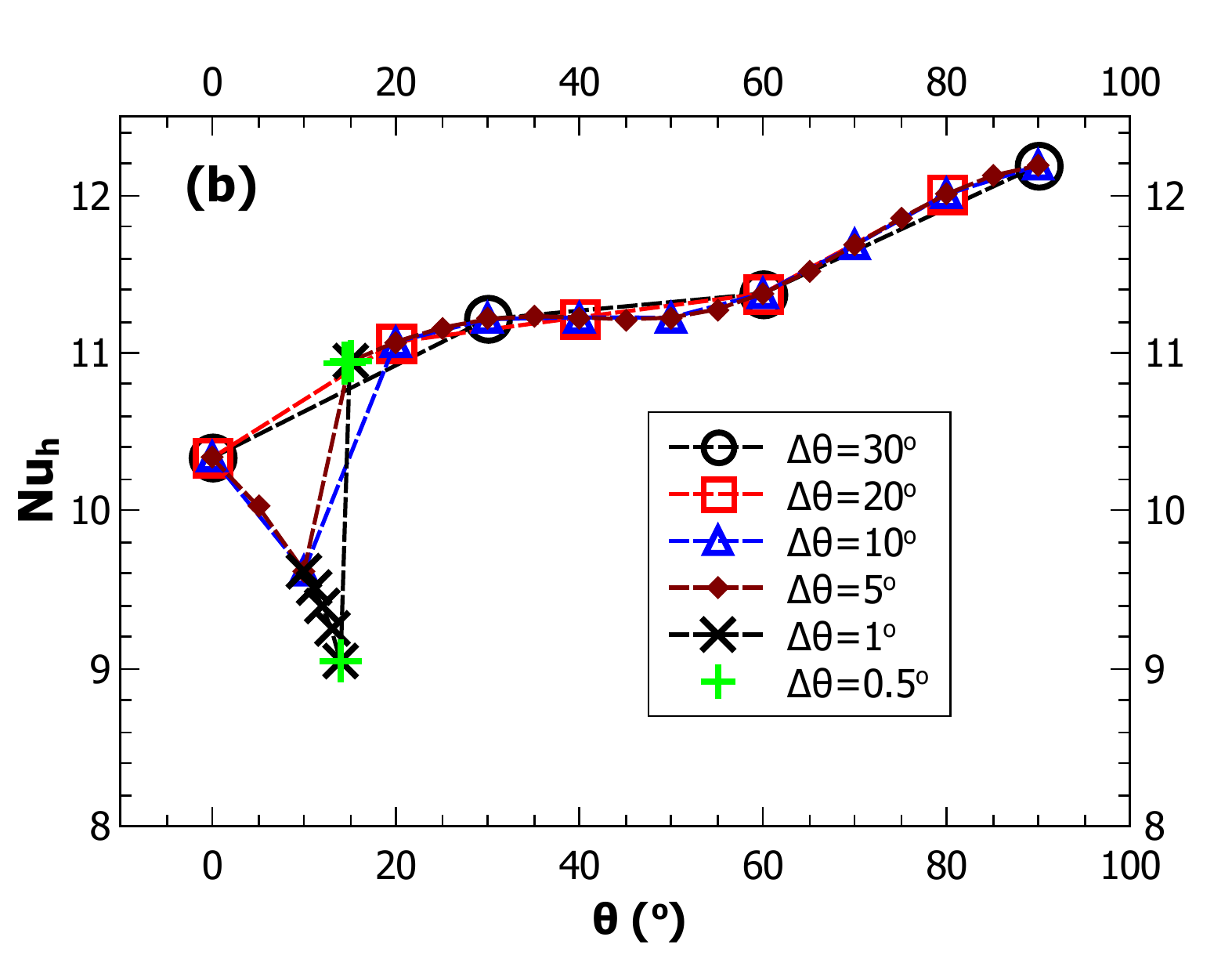}
	\end{subfigure}
	
	\caption{The effect of inclination on $Nu_h$: (a) different initial flow fields, (b) different inclination step sizes ($\Delta \theta$), for $Ra=1.6 \times 10^5$, $AR=1$, $D/H=0.4$, $Pr=0.7$ for $\theta: 0^\circ\; to\; 90^\circ$.}
	\label{fig:stepsizeeffectoninclinationnclystem}
\end{figure}

	\item Consideration of different axes for inclination study: For some of the studies reported in Table 1, different axis was considered for the inclination study. In our present study we consider the  NCL to be parallel to the $xy$ plane and the system is rotated along the $z$ axis to introduce the inclination. But some of the studies have considered $x$ axis for rotation, which does not induce any flow direction reversal and hence no heat transfer coefficient jump was observed. A discussion on the effect of inclination study w.r.t. the $x$ axis is presented in section.4.9.
\end{enumerate}

\subsection{Experimental validation}

Experimental work was conducted by Acosta et al. \cite{Acosta1987} with the primary focus of demonstrating the hysteresis effect and the existence of multiple steady states in the tilted NCL systems. The effect of inclination on the heat transfer coefficient was not reported or studied by them. From the present study it can be observed that the occurrence of the heat transfer coefficient jump in NCL systems is a consequence of the flow direction reversal resulting from the net buoyancy forces opposing the the fluid flow as discussed in sections 4.2 and 4.4.

Thus, if we recreate the experiment conducted by Acosta et al. \cite{Acosta1987} via a 3-D CFD study and observe a jump in the heat transfer coefficient at the point of flow direction reversal, then it can be confirmed that the heat transfer coefficient jump is a consequence of the flow direction reversal in NCL systems.

\begin{figure}[!htb]
	\centering
	\begin{subfigure}[b]{0.49\textwidth}
		\includegraphics[width=\linewidth]{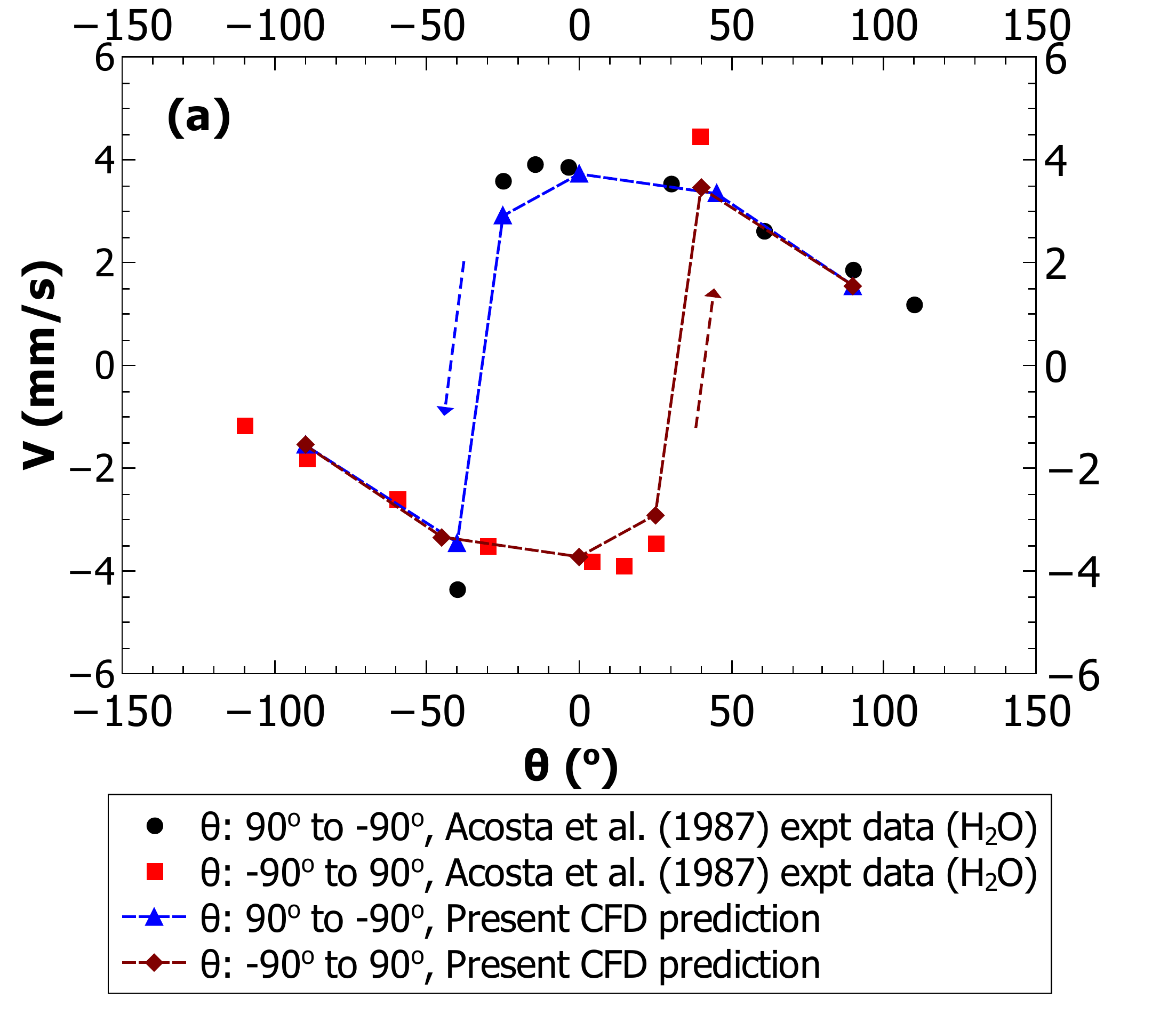}
	\end{subfigure}
	\hspace{\fill}
	\begin{subfigure}[b]{0.49\textwidth}
		\includegraphics[width=\linewidth]{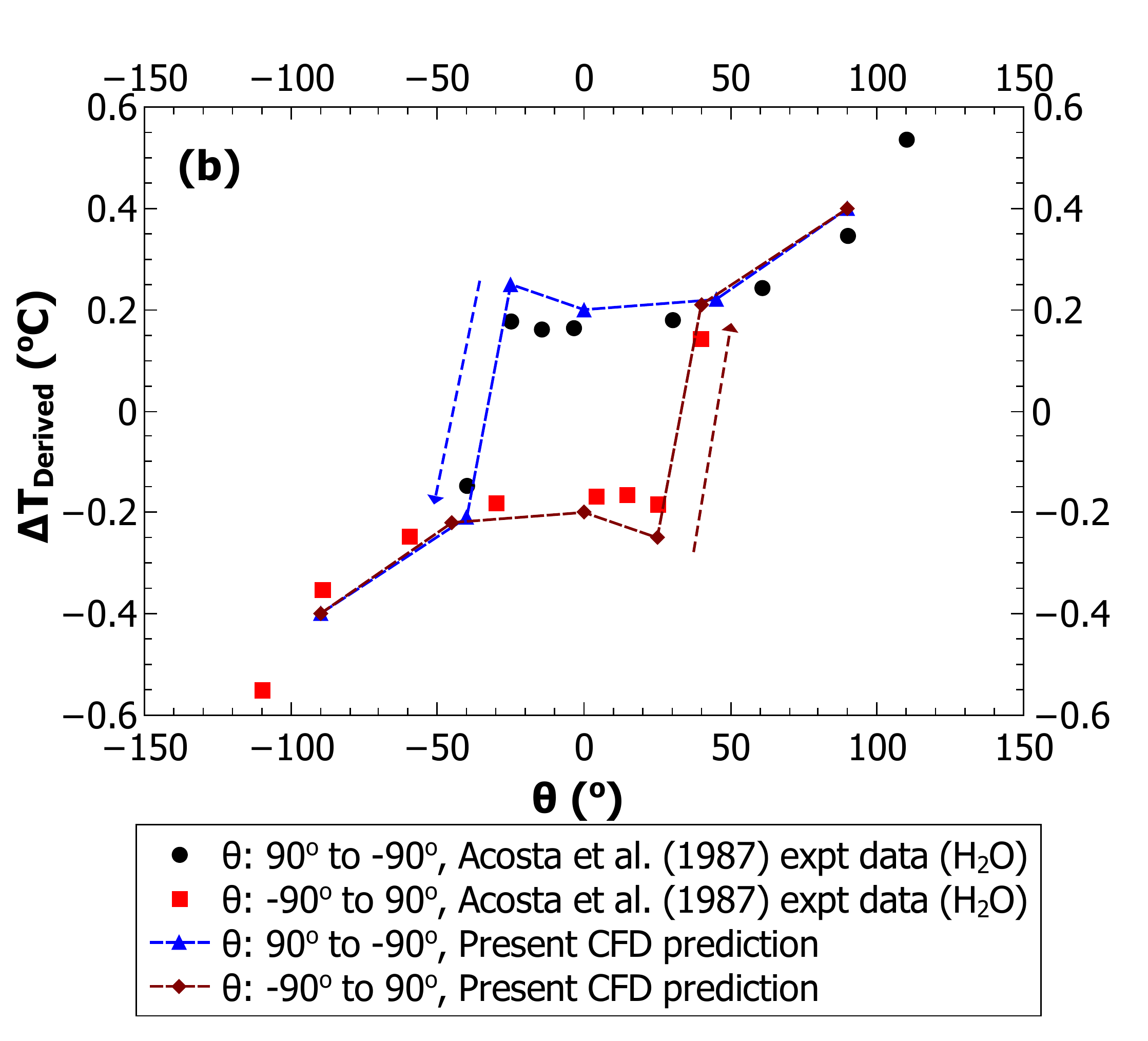}
	\end{subfigure}

	\caption{Validation of 3-D CFD with experimental data for a square NCL system: (a) Velocity ($V$) vs $\theta$,  (b) Temperature difference across the cooling section ($\Delta T_{Derived}$) vs $\theta$ with water as the working fluid for the case of $0.83 \;W$ heat transfer from the heated section \cite{Acosta1987}. }
	\label{Experimental validation}
\end{figure}

This validation also demonstrates other vital points such as:

\begin{enumerate}
	\item The use of 2-D CFD study does not restrict the physics of the inclination effects (the occurrence of the heat transfer coefficient jump) on the NCL systems.
	\item The working fluid used in the current 3-D CFD study is water and the working fluid considered in the previous sections is air ($Pr=0.7$), thus demonstrating the occurrence of the heat transfer coefficient jump being  independent of the fluid used.
	\item The validation of hysteresis behavior vis-a-vis heat transfer coefficient jump.
\end{enumerate}

The experimental set up of Acosta et al. \cite{Acosta1987} comprises of a square NCL system with the heat flux boundary condition on the bottom horizontal leg and a constant temperature boundary condition at the top horizontal leg. The equivalent geometry of the setup was created and a 3-D CFD study is performed. To obtain the steady state characteristics of the system the following intial assumptions are required for performing the 3-D CFD simulation: (a) The temperature at the cooling section is $50^\circ C$. A different temperature should not affect the solution considering the heat flux boundary condition at the heated section. (b) The initial temperature of the fluid is assumed to be the same as that of the cooling section. The 3-D CFD study is performed with the same case settings as described earlier for the 2-D CFD study. 

Figure \ref{Experimental validation} presents the validation of the 3-D CFD with the experimental data. A good match is observed between the CFD predictions and the experimental data indicating the accurateness of the present CFD study. Both velocity ($V$) and temperature difference across the heat exchanger ($\Delta T_{Derived}$) are used as the parameters for validation.

\begin{figure}[!htb]
	\centering
	\includegraphics[width=0.5\linewidth]{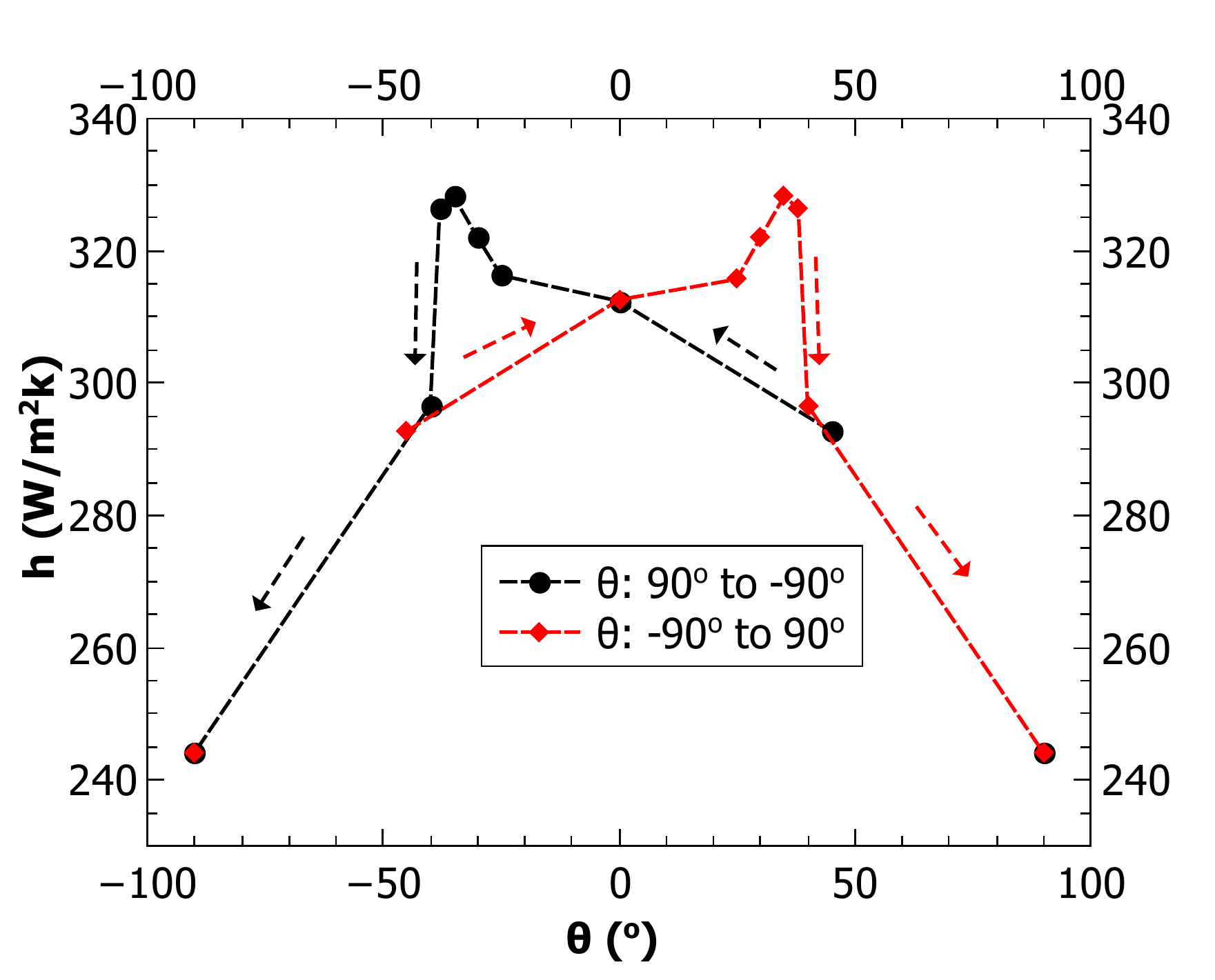}
	\caption{Effect of inclination on the heat transfer coefficient at the cooling section from the validated 3-D CFD study with water as the working fluid for the case $0.83 \;W$ heat transfer from the heated section \cite{Acosta1987}.}
	\label{fig:acostacfdh}
\end{figure}

Figure \ref{fig:acostacfdh} demonstrates the effect of the inclination on the heat transfer coefficient for the same case as in Figure 16. The heat transfer coefficient jump and the hysteresis behaviour w.r.t. the inclination in the considered NCL system can be clearly seen. Figure \ref{Experimental validation} and Figure \ref{fig:acostacfdh} together demonstrate that the heat transfer coefficient jump in the NCL systems is primarily because of the flow direction reversal.

To further understand the influence of different parameters on the heat transfer coefficient jump in the tilted NCL systems, a 2-D parametric study is presented in the subsequent sections.

\subsection{Effect of aspect ratio on the tilted NCL systems}

The aspect ratio of the NCL system may be defined w.r.t the height $H$ as follows:

\begin{equation}
AR=L/H
\end{equation}

%
%
%

\begin{figure}[!htb]
	\centering
	\begin{subfigure}[b]{0.49\textwidth}
		\includegraphics[width=1\linewidth]{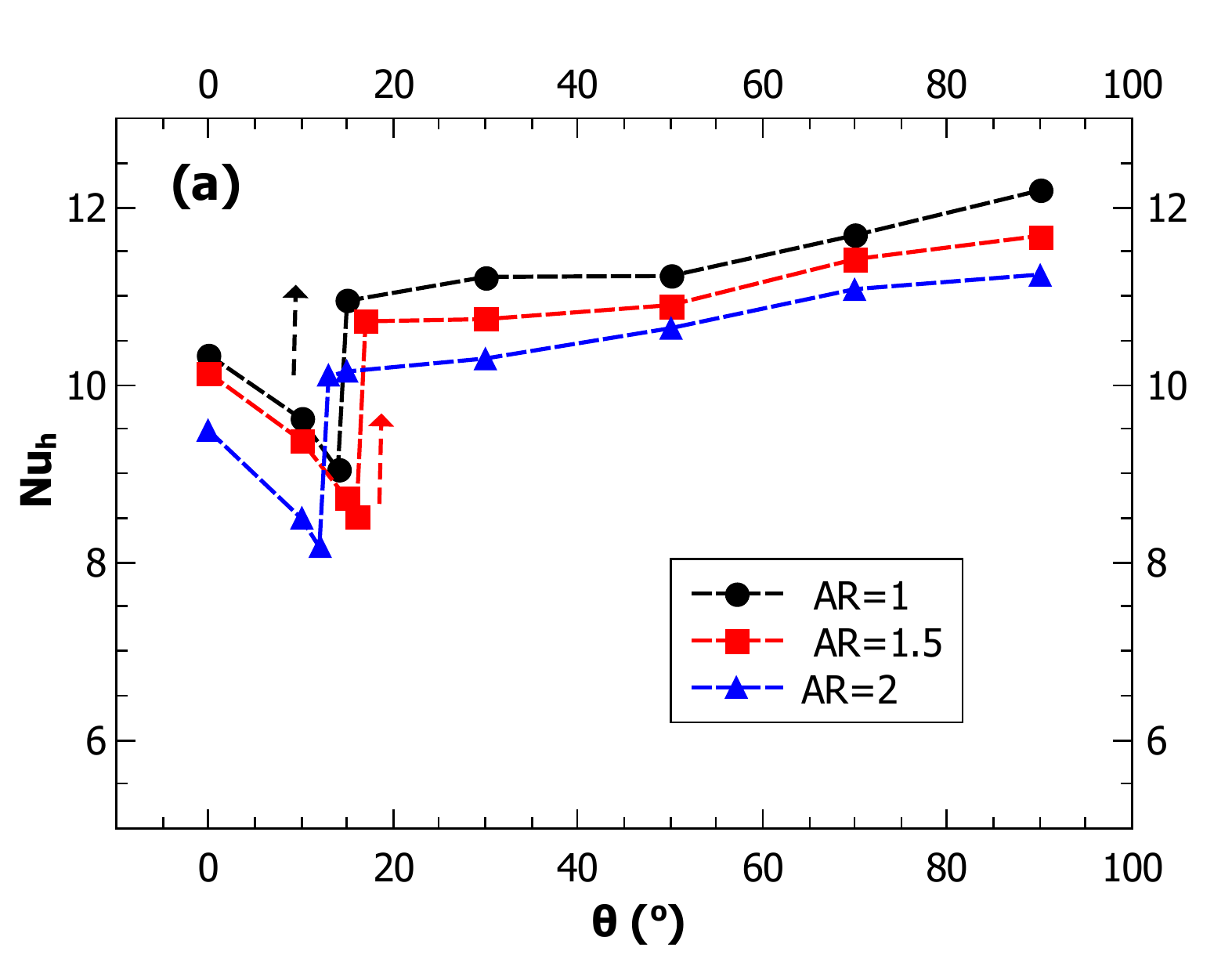}
	\end{subfigure}
	\hspace{\fill}
	\begin{subfigure}[b]{0.49\textwidth}
		\includegraphics[width=1\linewidth]{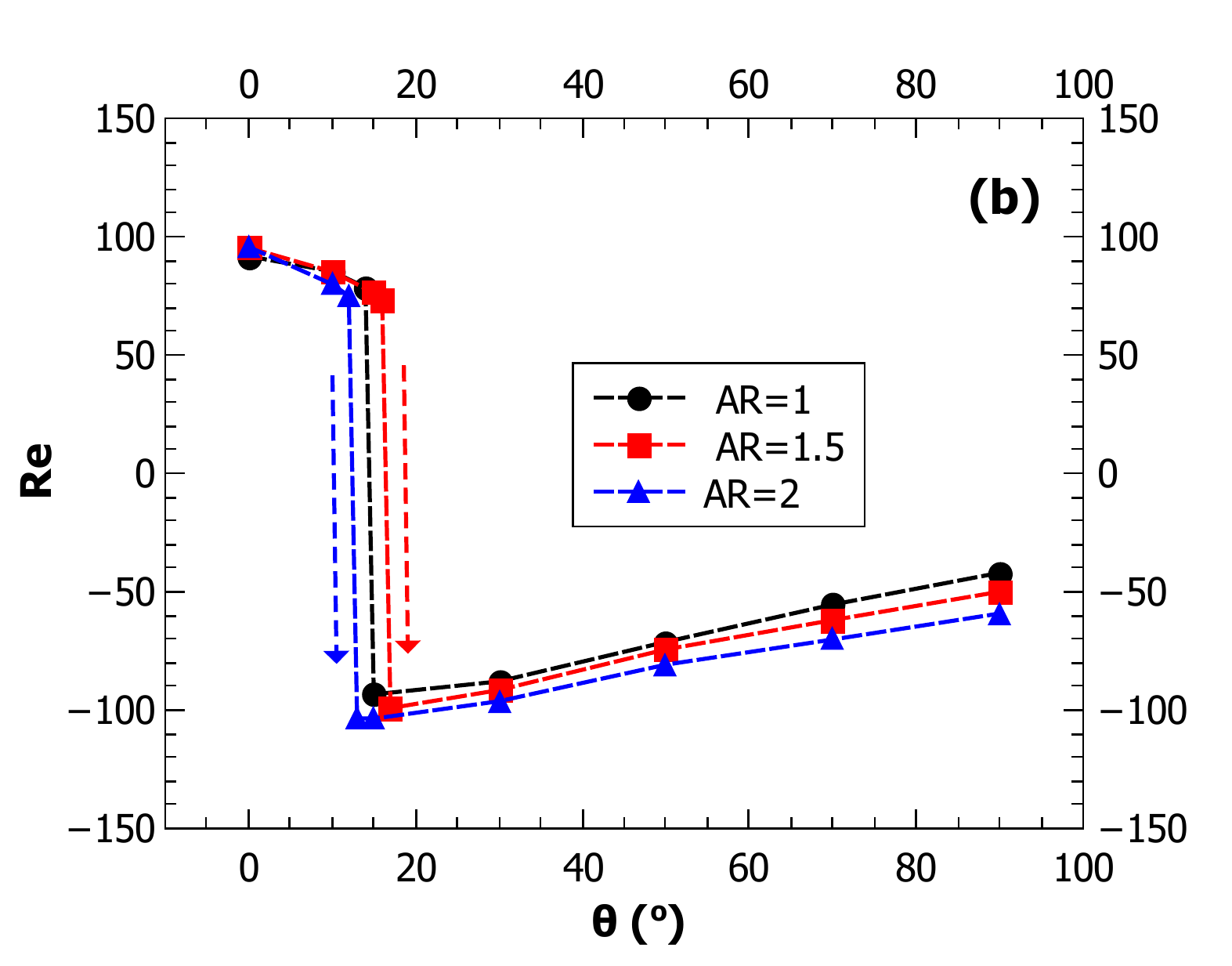}
	\end{subfigure}

	\caption{Effect of $AR$ on the the tilted NCL system: (a) $Nu_h$ and (b) $Re$, for $Ra=1.6 \times 10^5$, $D/H=0.4$ and $Pr=0.7$. }
	\label{AsH_inclinationNCL}
\end{figure}

Figure \ref{AsH_inclinationNCL} shows the effect of $AR$ on the variations of $Nu_h$ and $Re$ in the inclined NCL systems. It can be observed that the change in $AR$ does not shift significantly the point at which the heat transfer coefficient jump and the flow direction reversal occur. The influence of $AR$ on the NCL system can be understood by fixing the magnitude of $H$ to be unity for a particular fluid. Then, the variation of $AR$ amounts to the variation in  length ($L$) of the NCL system. By varying the $AR$ of the system the point at which the heat transfer coefficient jump occurs remains nearly unchanged as both the viscous and the inertial forces increase proportionally.

The net heat flux at the heated section of the wall ($q_w$) can be viewed as :

\begin{equation}
    q_w \propto k\frac{(T_h-T_c)}{H}
\end{equation}

As the magnitude of $H$ is kept fixed at unity, the heat flux remains the same for all the cases. This implies that the magnitude of buoyancy forces increases with increase in $L$, but as the viscous forces also increase with $L$ there appears to be no net effect of $AR$ on $Re$ and hence $Nu_h$. The observed differences in $Nu_h$ and $Re$ can be attributed to the possible change in the flow field at the corners of the NCL system. 

This explains why the $Nu_h$ and $Re$ versus $\theta$ are largely unaffected for all the considered aspect ratios.

\subsection{Effect of Rayleigh number on the tilted NCL systems}

\begin{figure}[!htb]
	\centering
	\begin{subfigure}[b]{0.49\textwidth}
		\includegraphics[width=1\linewidth]{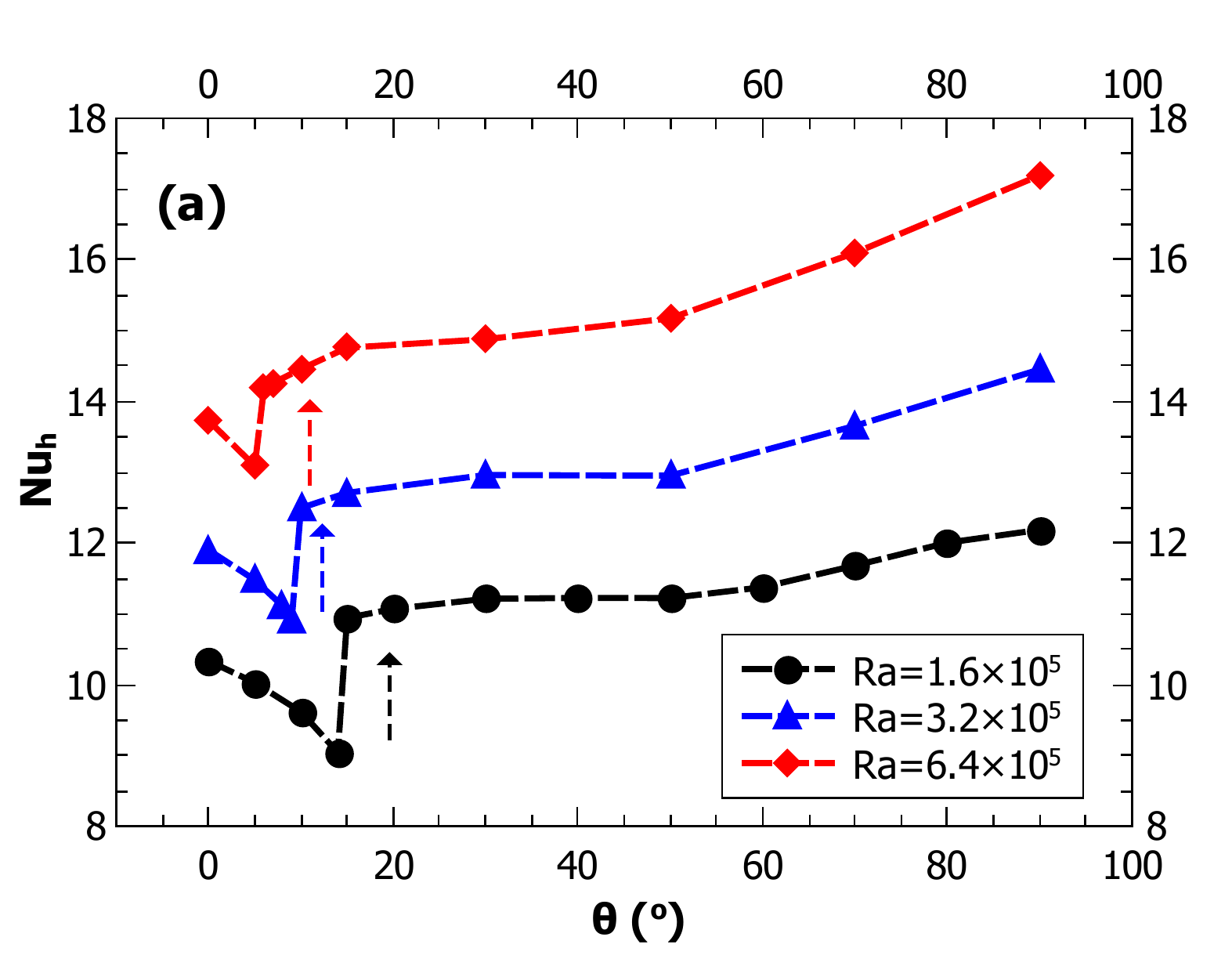}
	\end{subfigure}
	\hspace{\fill}
	\begin{subfigure}[b]{0.49\textwidth}
		\includegraphics[width=1\linewidth]{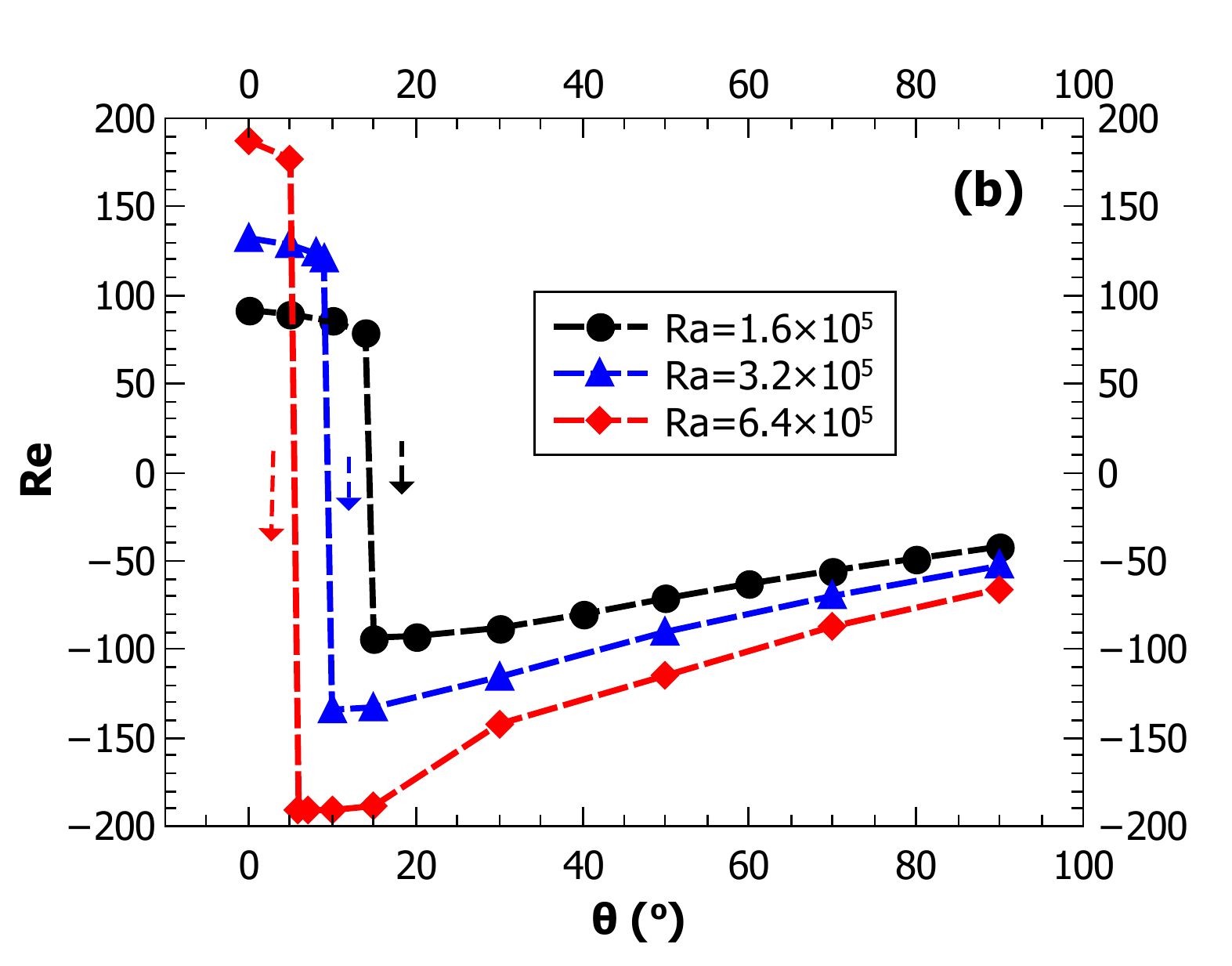}
	\end{subfigure}

	\caption{Effect of $Ra$ on the tilted NCL system: (a) $Nu_h$, (b) $Re$ for, $AR=1$, $D/H=0.4$ for $\theta: 0^\circ\; to\; 90^\circ$ for $Pr=0.7$.}
	\label{Ra effect on NCL}
\end{figure}

The effect of Rayleigh number on the $Nu_h$ and $Re$ of the tilted NCL systems is presented in Figure \ref{Ra effect on NCL}.
For a given $Ra$, the buoyancy forces are maximum for $\theta=0^\circ$ (heater at the bottom and cooler at the top), so the point flow reversal direction (the balance between the buoyancy force and the inertial force) approaches $\theta=0^\circ$ with increase in the $Ra$ value as shown in Figure \ref{Ra effect on NCL}(b). The same behaviour of the shift in  the point of flow-direction reversal has also been reported experimentally by Acosta et al. \cite{Acosta1987} with increase in the heat flux at the heated section of the NCL. The increase in the magnitude of $Ra$ is equivalent to the increase in the amount of energy supplied to a tilted NCL system. With increase in $Ra$, the buoyancy forces experienced by the system increases and this leads to an increase in the magnitude of $Re$, resulting in the higher $Nu_h$.

\subsection{Effect of the Prandtl number on the tilted NCL systems}

\begin{figure}[!htb]
	\centering
	\begin{subfigure}[b]{0.49\textwidth}
		\includegraphics[width=1\linewidth]{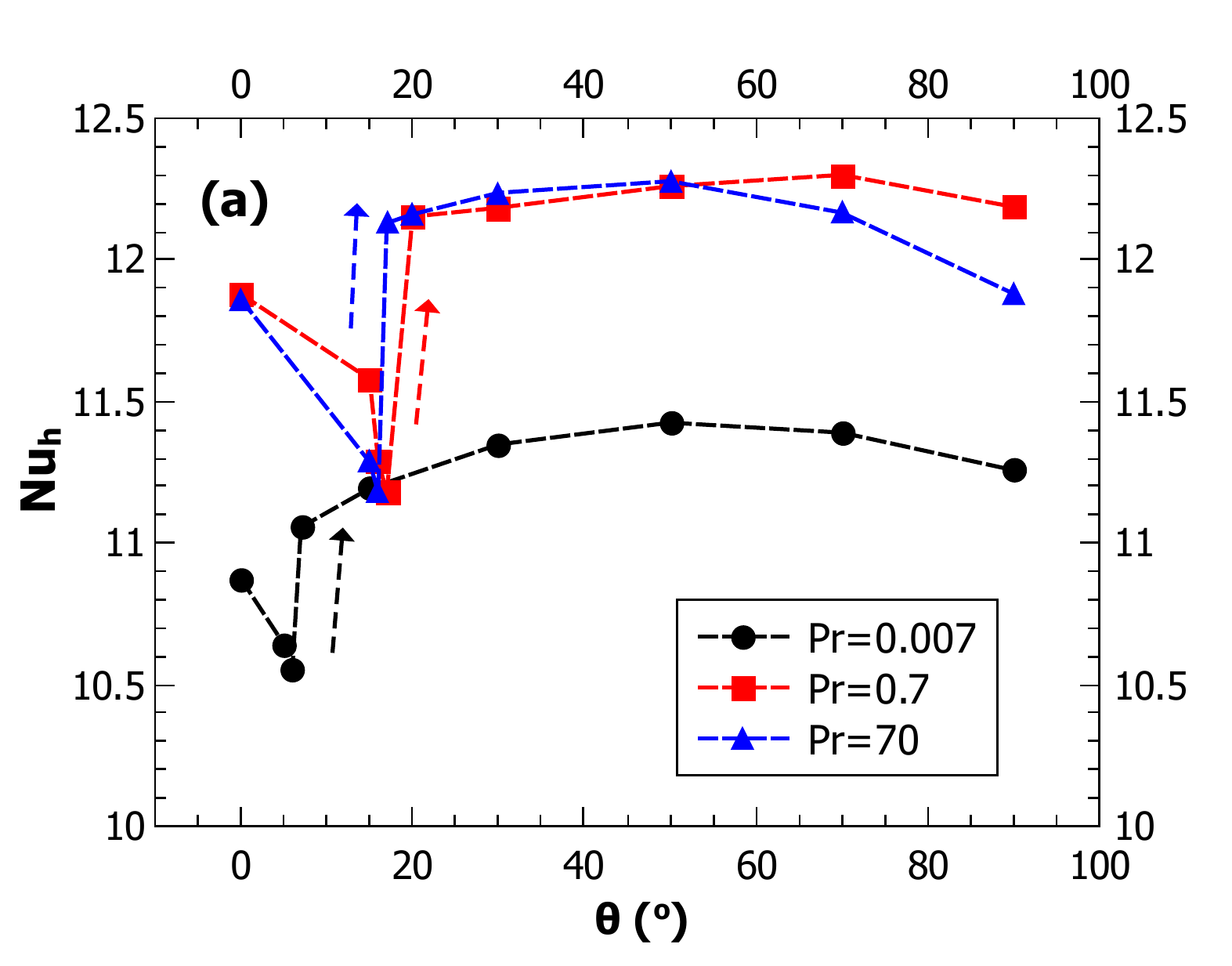}
	\end{subfigure}
	\hspace{\fill}
	\begin{subfigure}[b]{0.49\textwidth}
		\includegraphics[width=1\linewidth]{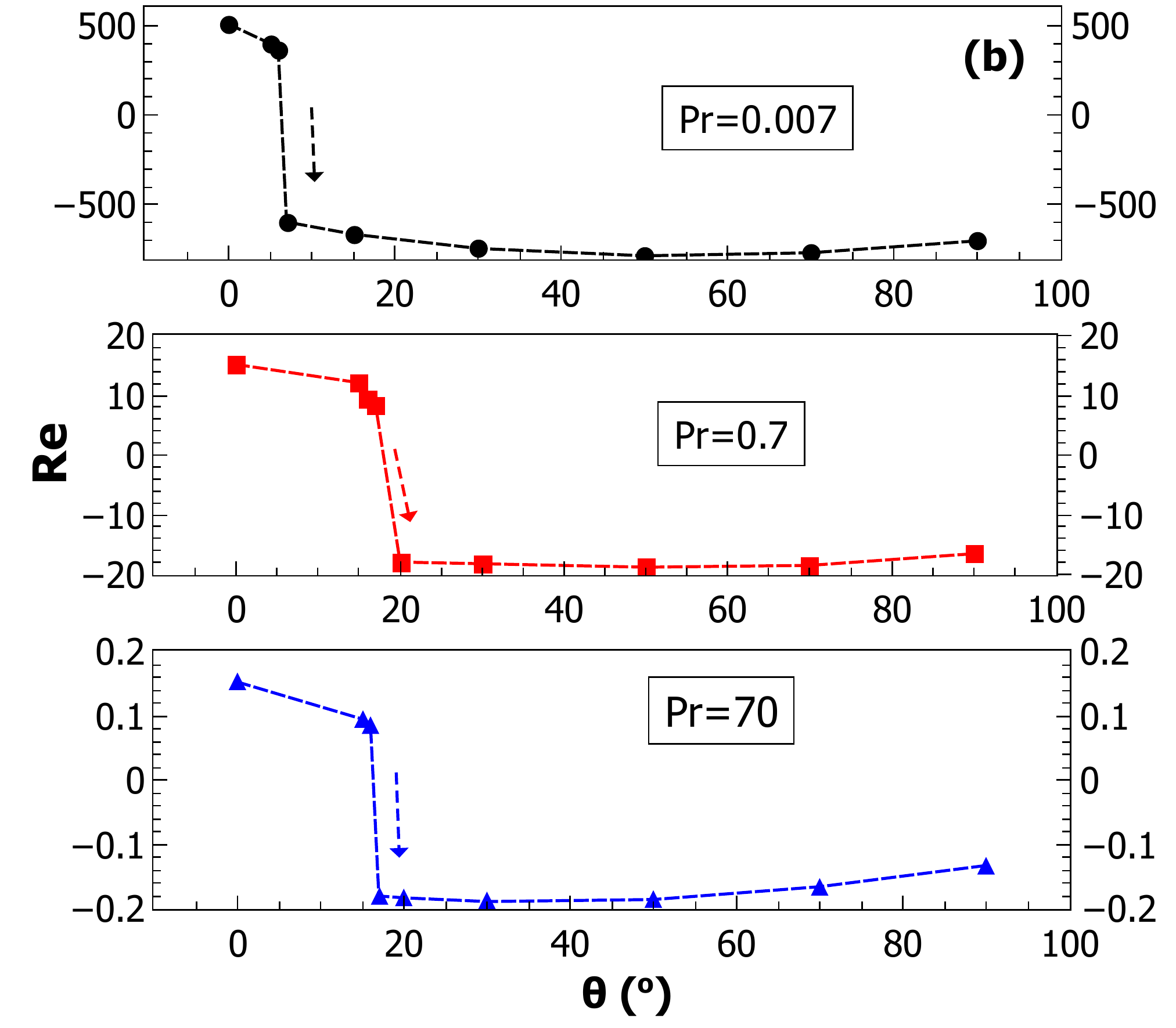}
	\end{subfigure}

	\caption{Effect of $Pr$ on the tilted NCL system: (a) $Nu_h$, (b) $Re$, $AR=1$, $D/H=0.4$ for $\theta: 0^\circ\; to\; 90^\circ$ for $Ra=16000$.}
	\label{Pr effect on NCL}
\end{figure}

The effect of variation of the Prandtl number on $Nu_h$ and $Re$ of the tilted NCL system is presented in Figure \ref{Pr effect on NCL} for a $Ra=16000$. The magnitude of chosen Prandtl numbers $0.007$, $0.7$ and $70$ correspond to liquid metals, air and viscous oils, respectively. Lower Prandtl number implies higher Grashof number or higher buoyancy forces for the same $Ra$ value. Hence the velocities or the $Re$ value is higher for lower Prandtl number as observed in Figure \ref{Pr effect on NCL}(b). Higher velocities coupled with higher thermal conductivities associated  with lower Prandtl number result in higher heat transfer coefficients. However, the Nusselt number being the ratio of the product of the heat transfer coefficient and the characteristic dimension to the thermal conductivity of the fluid, the $Nu_h$ appears to be almost same for all the three categories of the fluids ($Pr=0.007,0.7,70$) considered as shown in Figure \ref{Pr effect on NCL}(a). This is also in line with the dependence of the $Nu_h$ on the $Ra$ value which is kept constant for the case presented. The point of the occurrence of the jump approaches $\theta=0^\circ$ for the lower Prandtl number of $0.007$ due to higher buoyancy forces as explained in section.4.7.

\subsection{Effect of inclination $\gamma$ on the NCL system}

\begin{figure}[!htb]
	\centering
	\includegraphics[width=0.75\linewidth]{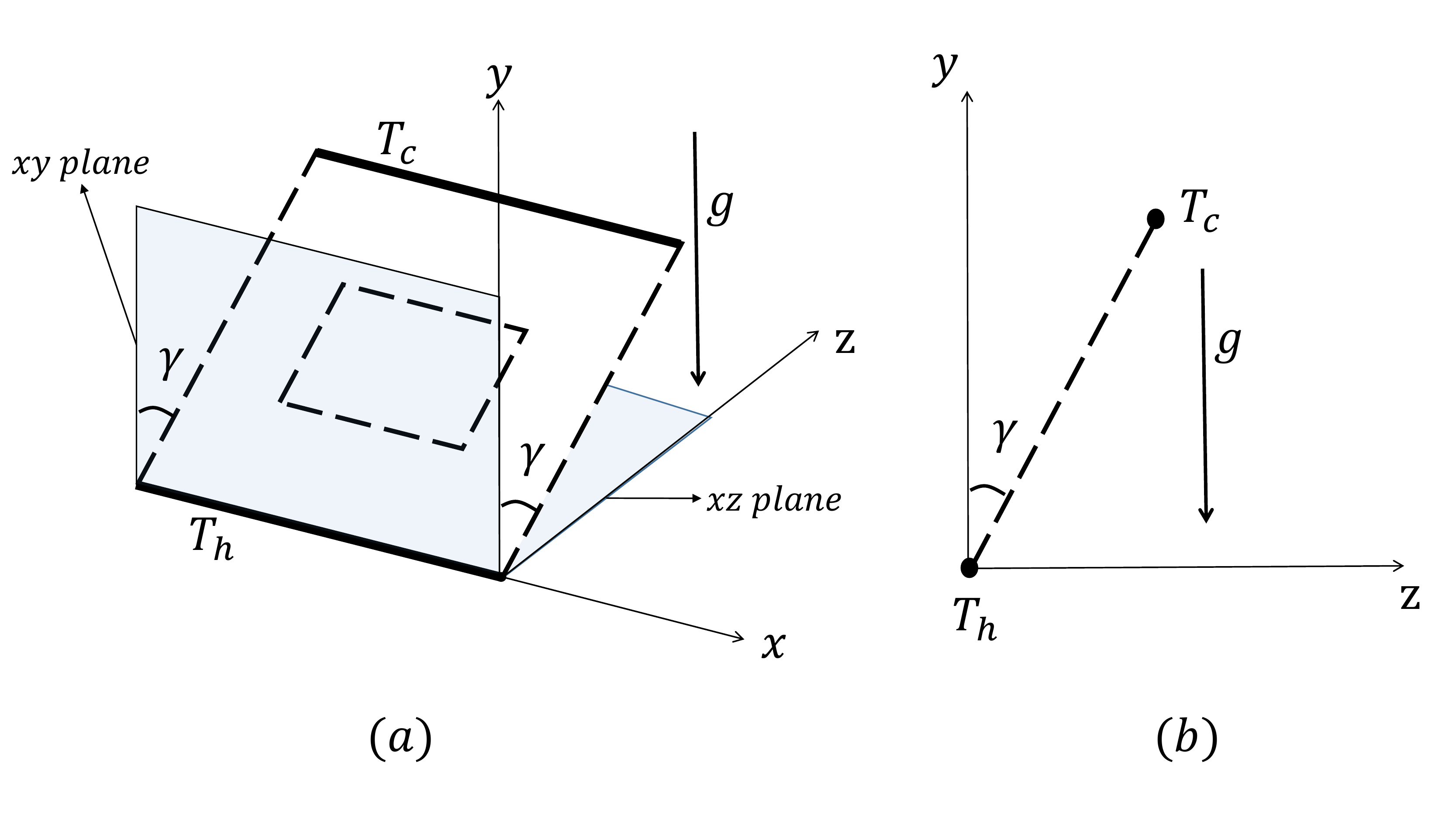}
	\caption{Schematic of the NCL system tilted by an angle $\gamma$ with respect to the $xy$ plane: (a) 3-D view of the inclination $\gamma$ w.r.t the xy plane, (b) Inclination $\gamma$ represented w.r.t the y-axis in the yz plane. The dashed lines are used to represent the insulated sections. }
	\label{fig:gammainclinationncl}
\end{figure}

\begin{figure}[!htb]
	\centering
	\includegraphics[width=0.5\linewidth]{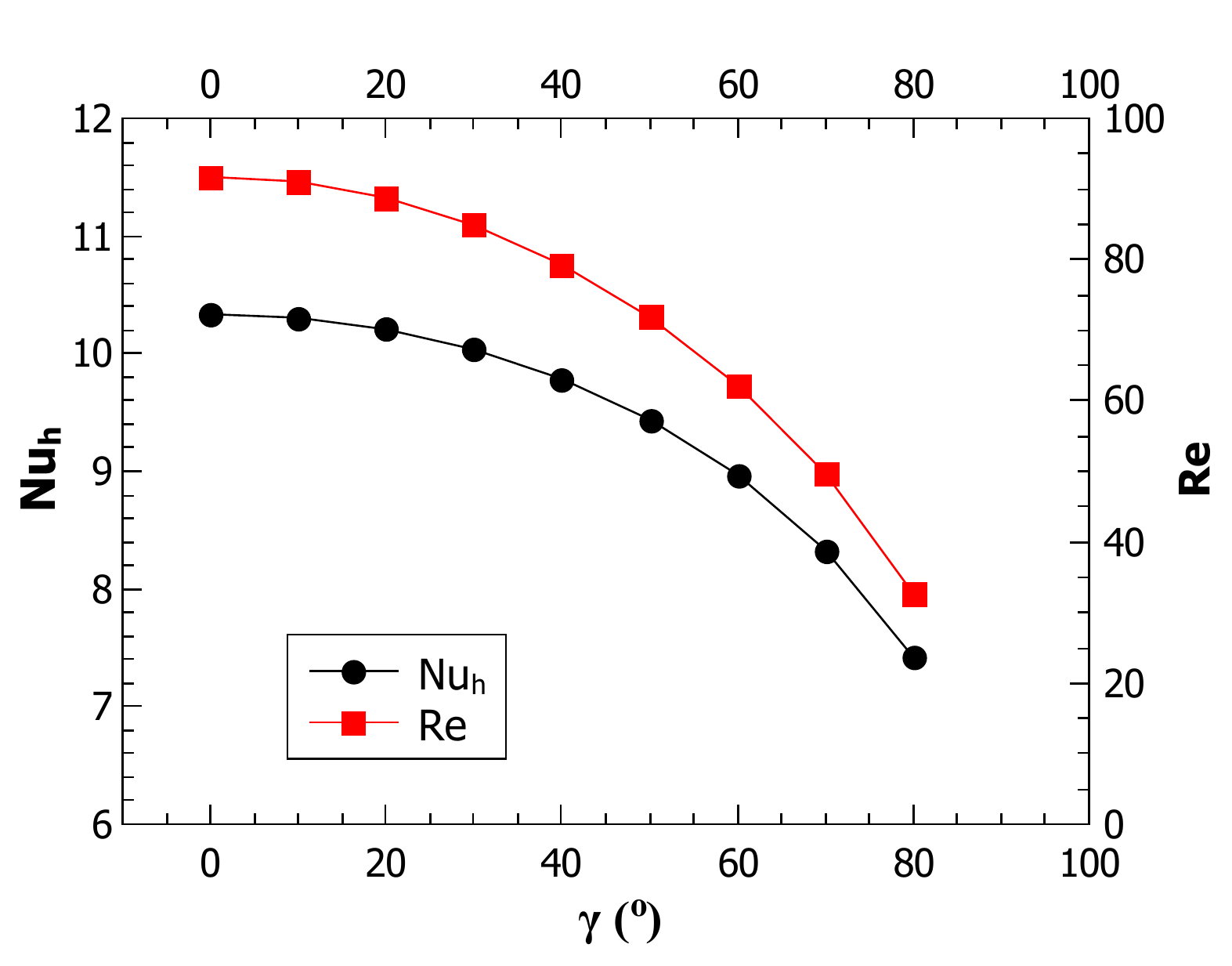}
	\caption{Effect of inclination $\gamma$ on the $Nu_h$ and $Re$ of the tilted NCL system for $Ra=1.6 \times 10^5$, $Pr=0.7$, $AR=1$, $D/H=0.4$. }
	\label{fig:Plotgammainclinationncl}
\end{figure}

 Figure \ref{fig:gammainclinationncl} illustrates the schematic of a tilted NCL system with respect to the $xy$ plane unlike the previous cases wherein the tilt is about the $z$ axis. The effect of inclination $\gamma$ on the NCL system is equivalent to reducing the body force experienced by the NCL system to $g \times cos(\gamma)$ and hence $Ra$ to $Ra \times cos(\gamma)$. This may be used to replicate the reduced gravity conditions as suggested by Garibaldi and Misale \cite{Garibaldi2008}. Thus, with the increase in the magnitude of $\gamma$ from $0^\circ$ to $90^\circ$, the magnitude of the body force decreases, which decreases the buoyancy forces, resulting in a reduction of both $Nu_h$ and $Re$ as shown in Figure \ref{fig:Plotgammainclinationncl}. The angle $\gamma=90^\circ$ corresponds to pure conduction. It can observed that there is no heat transfer coefficient jump, flow-direction reversal or hysteresis effect observed by the variation of angle $\gamma$. The introduction of angle $\gamma$ does not introduce any buoyancy forces which resist the fluid flow direction and hence no flow direction reversal is observed. Without the flow direction reversal no jump in the heat transfer coefficient is witnessed.  

\section{Conclusions}
In the current study  the effect of inclination on the buoyancy driven systems is presented. The present study indicates the occurence of the heat transfer coefficient jump and the hysteresis in tilted natural circulation systems such as NCL and CNCL, similar to cavity systems. The mechanism by which the heat transfer coefficient jump occurs in all the buoyancy driven systems is broadly a combination of the flow direction reversal and the flow pattern rearrangement. It is identified in  the present study that for a cavity of aspect ratio unity, NCL and CNCL systems, the flow direction reversal primarily dictates the point at which the heat transfer coefficient jump occurs. The conclusions from the present study can be summarised as follows:

\begin{enumerate}
    \item The present work thoroughly analyses as to why the occurrence of the heat transfer coefficient jump was not reported for NCL and CNCL systems. The study indicates that the step size has an important bearing on the observation of the heat transfer coefficient jump and the hysteresis. The study concludes that the initial condition used for the inclination effect influences the occurrence of the jump. If the steady state solution obtained for an inclination is used as the initial condition for the subsequent higher or lower angles corresponding to the increasing direction of $\theta$ or the decreasing direction of $\theta$, respectively, it results in the occurrence of the jump and the hysteresis. However, if the zero flow field (fluid at rest) initial condition is used for all $\theta$ then there occurs no jump and hysteresis.

    \item The 3-D CFD investigation, which has been validated with the available experimental data with water as the working fluid, predicts the occurrence of the heat transfer coefficient jump corresponding to the point of flow direction reversal.

    \item The magnitudes of $Nu_h$ and $Re$ of the tilted NCL systems are not significantly affected by the variation in $AR$ for the considered range. This is because of the proportional increase in the buoyancy forces and the viscous forces which balance each other. The point at which the heat transfer coefficient jump occurs is nearly independent of the variation in $AR$. 
    
    \item The increase in $Ra$ of the tilted NCL systems results in an increase in the magnitude of $Nu_h$ and $Re$. The point of jump in the heat transfer coefficient occurs earlier with increase in $Ra$.

    \item The heat transfer coefficient jump occurs for different fluids such as air and water and in general for all categories of fluids ranging from liquid metals to viscous oils.

    \item The inclination of the NCL plane w.r.t 'xy' plane does not induce any heat transfer coefficient jump or hysteresis phenomena unlike the NCL systems tilted about the $z$ axis.
    
    \item The authors have encountered oscillatory solutions for certain combination of parameters, which are not reported in the present study as they need an in-depth transient investigation and forms a scope for the future investigation. The work can be extended to different combinations of parameters involving $Pr$, $Ra$, $AR$, $D/H$ and heater-cooler positions. The way the experiments are conducted can influence the occurrence of the heat transfer coefficient jump as discussed in section 4.4, so the experimental investigation focused on the hysteresis effect and the heat transfer coefficient jump can be carried out.
   
\end{enumerate}

\section*{Acknowledgment}
The first author gratefully acknowledges the financial support in the form of doctoral scholarship received from Ministry of Human Resource Development (MHRD), Government of India.

\section*{\hfil \Large Nomenclature \hfil}

\begin{tabular}{ll}
    $A$ & Area of the cross section ($m^2$) \\
    $AR$ & Aspect ratio ($AR=L/H$)\\
    $C_p$ & Specific heat capacity of the fluid ($J/K$) \\
    $D_h$   & Hydraulic diameter of the NCL or CNCL system ($m$)\\
    $g$     & Gravitational acceleration constant ($m/s^2$)\\
	$h$ & Heat transfer coefficient ($W/m^2K$)\\
	$H$     & Height of the system ($m$)\\
	$k$ & Thermal conductivity ($W/mK$) \\
	$L$     & Width of the system ($m$)\\
	$p$     & Pressure ($Pa$)\\
	$P$     & Non-dimensional pressure \\
	$q_w$ & Heat flux at the wall ($W/m^2$) \\
	$Q^{\prime \prime}$ & Non-dimensional heat flux ($Q^{\prime \prime}=(q_w H)/k(T_h - T_c)$) \\
	
	$\Delta T_{Derived}$ & Temperature difference across the heat exchanger ($\Delta T_{Derived}=q_w/(\rho A C_p V_avg)$) \\
	$T_h$   & Hot wall temperature ($K$)\\
	$T_c$   & Cold wall temperature ($K$)\\
	$U$     & Non-dimensional x velocity \\
	$V$     & Non-dimensional y velocity \\
	$u$     &  x coordinate velocity ($m/s$) \\
	$v$     &  y coordinate velocity ($m/s$)\\
	$V_{avg}$ & Average velocity of the fluid in the NCL loop ($m/s$) \\
	$x$     & x coordinate ($m$)\\
	$y$     & y coordinate ($m$)\\
	$X$     & Non-dimensional x coordinate \\
	$Y$     & Non-dimensional y coordinate \\

    & \\
	\textbf{Non-dimensional numbers} & \\
	$Ra$    & Rayleigh number ($Ra=(g\beta (T_h - T_c) H^3)/\alpha \nu$)\\
	$Pr$    & Prandtl number ($Pr = \nu / \alpha$)\\
	$\phi$  & Non-dimensional temperature ($\phi = (T - T_0)/(T_h - T_c)$)\\
	$Nu$ & Nusselt number ($Nu=(hH)/k$)\\
	$Re$ & Reynolds number ($Re=(V_{avg} H)/\mu $) \\
	& \\
	\textbf{Greek letters} & \\
	$\rho$ & Density ($Kg/m^3$) \\
	$\alpha$ & Thermal diffusivity ($m^2/s$)\\  
    $\beta$ & Thermal expansion coefficient ($1/K$)\\
    $\nu$ & Kinematic viscosity  ($m^2/s$)\\
    $\mu$ & Dynamic viscosity  ($kg/ms$)\\
    $\theta$& Inclination of NCL system about $z$ axis ($^\circ$)\\
	$\gamma$ & Inclination of NCL plane w.r.t. the $xy$ plane ($^\circ$)\\

\end{tabular}

\begin{tabular}{ll}
	\textbf{Abbreviations} & \\
	$CFD$ & Computational Fluid Dynamics \\
	$NCL$ & Natural Circulation Loop \\
	$CNCL$ & Coupled Natural Circulation Loop \\
	$PRHRS$ & Passive Residual Heat Removal Systems \\
	& \\
	\textbf{Subscripts} & \\
	$0$ & Any parameter considered at initial condition\\
	$1$ & Any parameter considered for Loop 1 of the CNCL \\
	$h$ & Any parameter considered at the heated section \\
	$CHX$ & Any parameter considered at the common heat exchange section of CNCL system \\
	
\end{tabular}

\bibliography{ICHMT}

\end{document}